%% file: conference_101719.tex
\documentclass[conference]{IEEEtran}
\IEEEoverridecommandlockouts
\usepackage{cite}
\usepackage{amsmath,amssymb,amsfonts}
\usepackage{algorithmic}
\usepackage{graphicx}
\usepackage{textcomp}
\usepackage{xcolor}
\usepackage{color}
\usepackage{subcaption}
\usepackage{enumitem}
\usepackage{pifont}
\usepackage{caption}
\usepackage{threeparttable}
\usepackage{makecell}
\usepackage[bookmarks=true,breaklinks=true,letterpaper=true,colorlinks,citecolor=blue,linkcolor=blue,urlcolor=blue]{hyperref}
\usepackage[mathscr]{eucal}
\usepackage[ruled,linesnumbered]{algorithm2e}
\usepackage{booktabs}
\usepackage{ragged2e} 
\usepackage{balance}
\usepackage{microtype}
\usepackage{booktabs,makecell, multirow, tabularx}
\usepackage{color}
\usepackage{enumitem}

\usepackage{hyperref}
\hypersetup{
    colorlinks=true,  % 设置链接颜色为文本颜色，而不是边框颜色
    linkcolor=black,  % 设置内部链接颜色为黑色
    urlcolor=black,   % 设置 URL 链接颜色为黑色
    citecolor=black,  % 设置引用链接颜色为黑色
    linkbordercolor=white,  % 设置链接边框颜色为白色，即无边框
    urlbordercolor=white,   % 设置 URL 边框颜色为白色，即无边框
    citebordercolor=white,  % 设置引用边框颜色为白色，即无边框
    pdfborderstyle={/S/U/W 0}  % 设置边框样式为无边框
}

\def\BibTeX{{\rm B\kern-.05em{\sc i\kern-.025em b}\kern-.08em
    T\kern-.1667em\lower.7ex\hbox{E}\kern-.125emX}}

\newcommand{\etal}{\emph{et al.}}
\newcommand{\xcut}{X-cut}
\newcommand{\zcut}{Z-cut}
\newcommand{\cnot}{CNOT}
\newtheorem{problem}{\text{Problem}}
\newcommand{\problemtitle}[1]{\textbf{#1}}
\newtheorem{theorem}{Theorem}
\newtheorem{lemma}{Lemma}
\newtheorem{definition}{\text{Definition}}
\newcommand{\para}{Circuit Parallelism Degree}
\newcommand{\chipcapacity}{Chip Communication Capacity}
\newcommand{\PRX}{EDPCI}
\newcommand{\unlimit}{ReSu}
\newcommand{\ourframework}{\texttt{Ecmas}}
\usepackage[english]{babel}

\begin{document}

% \title{An efficient Resource-Adaptive Compilation Framework for Surface Code Quantum Computing}
\title{Ecmas: Efficient Circuit Mapping and Scheduling for Surface Code}

\author{
    \IEEEauthorblockN{
        Mingzheng Zhu\IEEEauthorrefmark{2}, 
        Hao Fu\IEEEauthorrefmark{2},
        Jun Wu\IEEEauthorrefmark{2},
        Chi Zhang\IEEEauthorrefmark{2},
        Wei Xie\IEEEauthorrefmark{2},
        Xiang-Yang Li\IEEEauthorrefmark{2}\IEEEauthorrefmark{3}
    }
    \IEEEauthorrefmark{2}{
        University of Science and Technology of China, China
    }\\
     \IEEEauthorrefmark{3}{
        Hefei National Laboratory, University of Science and Technology of China, Hefei 230088, China
    }
}
\maketitle

\begin{abstract}
\input{0-Abstract}
\end{abstract}
\begin{IEEEkeywords}
Surface Code, Compilation, Execution Time
\end{IEEEkeywords}
\input{1-Introduction}
\input{2-Background}
\input{3-ProblemFormulation}

\input{5-SysDesign}
\input{6-Evaluation}
\input{7-RelatedWork}

\input{8-Conclusion}
\section*{Acknowledgements}
The research is partially supported by National Key R\&D Program of China under Grant No.2021ZD0110400, Innovation Program for Quantum Science and Technology 2021ZD0302901, Anhui Initiative in Quantum Information Technologies under grant No. AHY150300 and China National Natural Science Foundation with No. 62132018, "Pioneer" and "Leading Goose" R\&D Program of Zhejiang", 2023C01029, and 2023C01143. Xiang-Yang Li is the corresponding author (Email: xiangyangli@ustc.edu.cn).

% \appendices
\begin{appendices}
\input{A}
\end{appendices}

\balance

\bibliographystyle{IEEEtranS}
\bibliography{refs}

\end{document}

%% file: 0-Abstract.tex
As the leading candidate of quantum error correction codes, surface code suffers from significant overhead, such as execution time. Reducing the circuit's execution time not only enhances its execution efficiency but also improves fidelity. However, finding the shortest execution time is NP-hard.

In this work, we study the surface code mapping and scheduling problem. To reduce the execution time of a quantum circuit, we first introduce two novel metrics: \para~and \chipcapacity~to quantitatively characterize quantum circuits and chips. Then, we propose a resource-adaptive mapping and scheduling method, named \textbf{\ourframework}, with customized initialization of chip resources for each circuit. \textbf{\ourframework}~can dramatically reduce the execution time in both double defect and lattice surgery models. Furthermore, we provide an additional version \ourframework-\unlimit~for sufficient qubits, which is 
performance-guaranteed and
more efficient.
Extensive numerical tests on practical datasets show that \ourframework~outperforms the state-of-the-art methods by reducing the execution time by 51.5\% on average for double defect model. \ourframework~can reach the optimal result in most benchmarks, reducing the execution time by up to 13.9\% for lattice surgery model.

%我们展示了一种普遍有效的方法，更具线路的需求提供芯片的资源。

%降低线路的执行时间，不仅能够增加线路的执行效率，并且也能增强线路的保真度。

%% file: 1-Introduction.tex
\section{Introduction}
Quantum algorithms offer exponential speedup over classical algorithms in various fields such as machine learning \cite{huang2021power,schuld2019quantum,havlivcek2019supervised}, simulation \cite{georgescu2014quantum} and cryptography \cite{shor1994algorithms}. 
One of the obstacles to achieving such advantages is the inevitable errors of quantum hardware.
The error rate of the state-of-the-art superconducting quantum devices is around $10^{-3}$ per operation \cite{wu2021strong,arute2019quantum,grover1996fast}, which falls far short of meeting the demands of practical applications \cite{gidney2021factor}. 
One approach to handle these errors is quantum error correction (QEC), which establishes the fault-tolerant computational framework \cite{shor1996fault}.
Surface code \cite{bravyi1998quantum, dennis2002topological, kitaev2003fault} currently stands as the most promising QEC code, highlighting a threshold error rate of up to $10^{-2}$. Its natural 2-D nearest-neighbor structure makes it well-suited for implementation on superconducting chips.

Applying surface code to protect a quantum circuit involves converting the circuit into an encoded form. Unlike the circuit transformation typical in the NISQ (Noisy Intermediate-Scale Quantum) era, surface code transformation necessitates mapping a single logical qubit to a cluster of physical qubits, known as tiles\cite{javadi2017optimized}. As a result, the conditions for executing logical operations differ significantly. For instance, a CNOT gate no longer requires physical qubits to be adjacent on the chip. Instead, it can be implemented by establishing a non-intersecting path between two distinct tiles, called qubit communication. These requirements call for developing an efficient and specialized transformer to transform a quantum circuit into a surface-code-encoded circuit.

The transformation process has two stages: initialization and scheduling. In the initialization stage, the transformer needs to allocate tiles for each logical qubit and allocate channels for communication. In the scheduling stage, the transformer determines the specific execution schemes for each operation. Most operations can be performed within the tiles \cite{fowler2012surface}, except T gate and CNOT gate, which are the most resource-consuming logical operations.
The substantial overhead of T gates stems from their inability to be fault-tolerantly executed, thereby necessitating the use of supplementary \emph{magic state distillation} circuits \cite{bravyi20magic12}. Through extensive research efforts \cite{ding2018magic,litinski2019game}, this overhead has been considerably reduced. 
However, the time delay induced by CNOT gates is severe, particularly in circuits where quite a lot of CNOT gates can be executed in parallel, such as Ising circuits \cite{javadiabhari2014scaffcc,Qiskit} and the QDNN circuits \cite{stein2022quclassi}. This significantly influences the fidelity of the execution result of the surface code circuit. With the same physical error rates and the code distance, a shorter execution time yields higher result fidelity. Thus, an essential goal of transformation is to reduce the execution time of the circuit. 
However, finding the shortest execution time of a circuit is NP-hard (as proved in Theorem \ref{hardness}), which makes it a non-trivial task for finding an effective result efficiently.

\begin{figure}[htbp]
	\centering  %图片全局居中
	\includegraphics[width=0.9\linewidth]{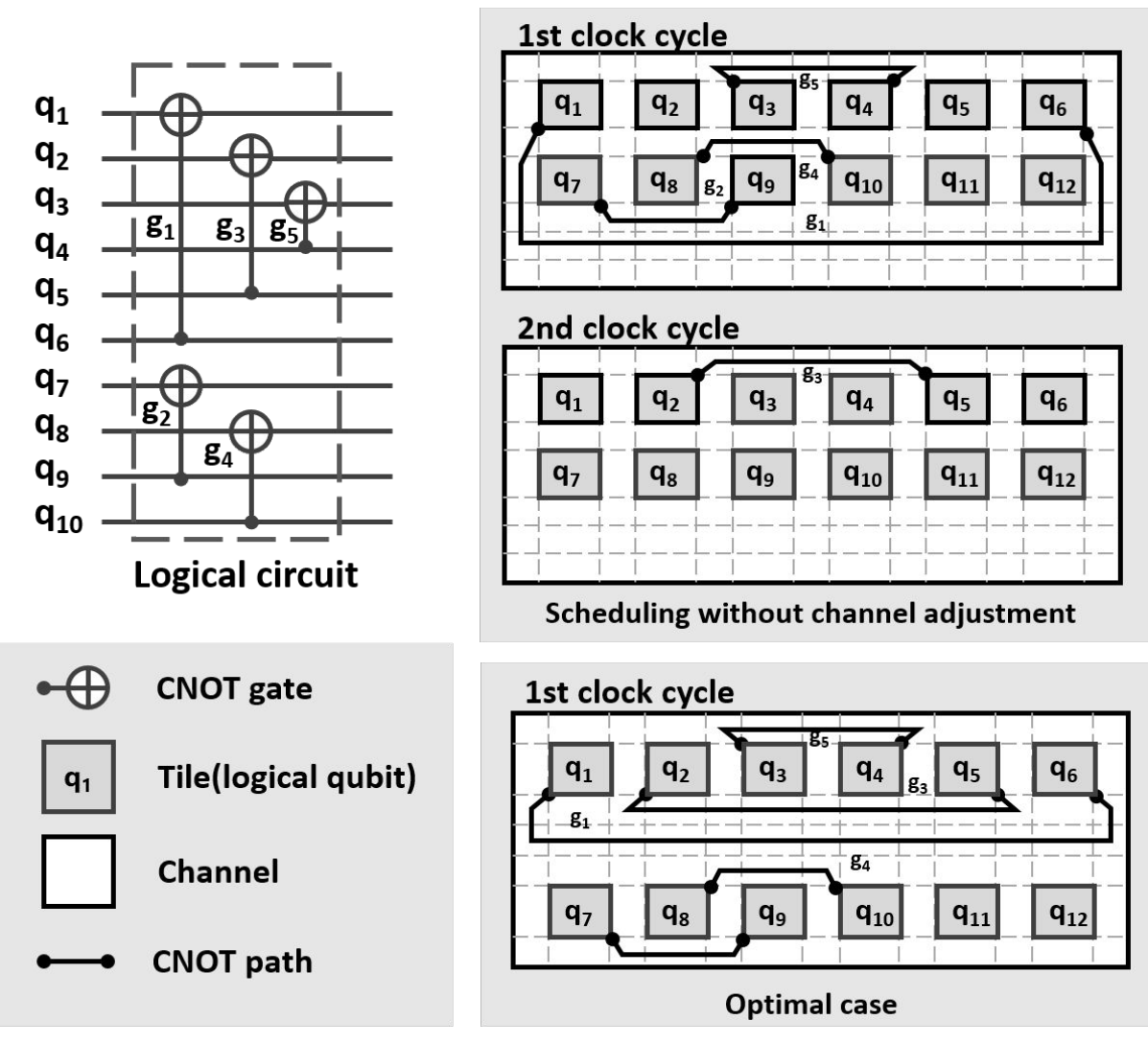}
	\caption{Motivation example: a logical quantum circuit and its corresponding surface code encoded circuit. }
	\label{motivation}
\end{figure}

Executing CNOT gates can be simplified as constructing a path between the two involved tiles, regardless of the specific encoding scheme, i.e., double defect model \cite{fowler2012surface} or lattice surgery model \cite{beverland2022surface}. Logical qubits are represented as small boxes in Fig. \ref{motivation}, and channels are the residual regions used to establish the paths (depicted by the lines). CNOT gates can be completed within one clock cycle, regardless of the path length. 
Simultaneous execution of CNOT gates requires non-intersect corresponding paths.

Many works focus on the paths of CNOT gates to reduce the latency caused by path conflicts \cite{javadi2017optimized,hua2021autobraid,beverland2022surface}. 
Braidflash \cite{javadi2017optimized} reduces the latency of CNOT gates on the critical path by assigning priority to CNOT gates to reduce the delay of the conflicts.
AutoBraid \cite{hua2021autobraid} identifies specific patterns to find non-intersecting paths and \PRX~\cite{beverland2022surface} draws inspiration from the concept of edge-disjoint paths. However, they have all overlooked a crucial aspect: within the context of surface code, the communication resources on the chip are software-defined. We use \textbf{bandwidth} to represent the width of the channel, with which we can adaptively adjust communication resources for varying circuits.

\noindent\textbf{Motivation Example:} As shown in Fig. \ref{motivation}, five independent gates are ready to be executed. Each gate's path requires a certain width of physical qubits during execution. Due to the lack of quantitative analysis on the channels, the path occupies the entire channel in the scheduling process of previous works. As a result, no five disjoint paths allow these gates to be executed simultaneously. However, in the optimal case, one clock cycle is enough to execute the circuit with the same chip and tile placement by better allocating the channel resource.

In this paper, we study the surface code circuit mapping and scheduling problem. We propose resource-adaptive transforming methods \ourframework~which reduce the execution time of circuits on target chips by customizing channel resources for each circuit. The main idea is to characterize the circuit and the chip and then schedule communication resources based on the circuit's requirements. Our contributions are summarized as follows:
\begin{itemize}
    \item We formulate the surface code circuit mapping and scheduling problem and analyze the computational complexity of double defect model.
    \item We define \textit{\para}~and \textit{\chipcapacity}~to quantitatively characterize circuits and chips. Further, we introduce \textit{bandwidth} to customizing channel resources for each quantum circuit.
    \item We propose resource-adaptive mapping and scheduling methods that can be applied to both double defect and lattice surgery models. With sufficient physical qubits, \ourframework~offers \ourframework-\unlimit~which can provide performance-guaranteed results efficiently.
    \item We evaluate \ourframework~for circuits from IBM Qiskit \cite{Qiskit}, QASMbench \cite{li2022qasmbench}, etc. With the same chip resource configuration, \ourframework~eliminates 51.5\% of the execution time on average and 67.3\% at most compared with Autobraid \cite{hua2021autobraid} for double defect model. \ourframework~could find the optimal result for most benchmarks for lattice surgery model. Compared with \PRX~\cite{beverland2022surface}, \ourframework~can achieve optimizations up to 13.9\%.
\end{itemize}

The rest of this paper is organized as follows. We introduce the background in Section \ref{background} and then formulate the surface code mapping and scheduling problem in Section \ref{problem_hard}. We describe our methods in Section \ref{sec:system} and evaluate their performance in Section \ref{sec:experiment}. Related works and conclusion are given in Section \ref{sec:related work} and Section \ref{sec:conclusion}.

%% file: 2-Background.tex
\section{Background}             

In this section, we present a brief overview of quantum error correction and surface code model.
\label{background}
\subsection{Quantum Error Correction}

Quantum programs can be described by the quantum circuit model, which consists of a sequence of quantum gates performed upon a collection of qubits. Qubits are the fundamental units in quantum computing which can be represented by a normalized vector. Quantum gates are unitary operations that operate on qubits.

However, quantum computing suffers from the inevitable noise of interactions with the environment and imprecise operations. QEC codes are necessary to build fault-tolerant quantum computing. It encodes a logical qubit with multiple physical qubits, improving reliability. The noise of the quantum system appears not only in the communication process but also in the computation process. Therefore, the quantum circuit must run under the protection of QEC. QEC codes should detect and correct errors periodically during the execution.

\subsection{Surface Code}

Among various QEC codes, surface code is a prominent candidate for achieving fault-tolerant quantum computation in superconducting implementations. It has a high threshold of around 1\% and alignment with 2D architectures, making it a feasible error correction code for practical demonstrations on real machines \cite{google2023suppressing,krinner2022realizing,zhao2022realization}.  

\begin{figure}[htbp]
    \centering
    \begin{subfigure}[b]{0.4\linewidth}
        \includegraphics[width=\linewidth]{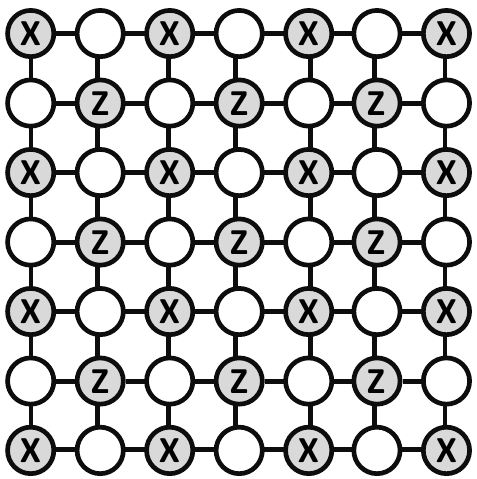}
        \subcaption{}
        \label{surface_code}
    \end{subfigure}
    \begin{subfigure}[b]{0.22\linewidth}
        \includegraphics[width=\linewidth]{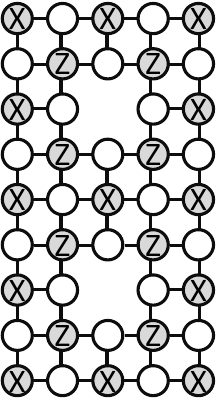}
        \subcaption{}
        \label{x-mea}
    \end{subfigure}
        \begin{subfigure}[b]{0.21\linewidth}
        \includegraphics[width=\linewidth]{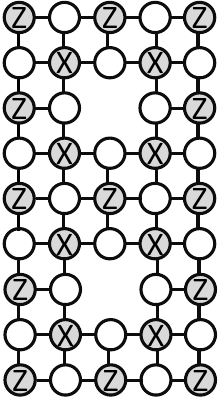}
        \subcaption{}
        \label{z-mea}
    \end{subfigure}
    \caption{(a) Surface code implementation on 2-D lattice, the white circles are data qubits and the gray circles are measurement qubits, (b) X-cut tile, (c) Z-cut tile.}
	\label{surface_code_intro}
	\vspace{-0.2cm}
\end{figure}

As shown in Fig. \ref{surface_code_intro}, surface code is realized on a 2D lattice of physical qubits, including data and measurement qubits. Data qubits store quantum states, while measurement qubits identify error occurrences. Based on the measurement circuit, measurement qubits are categorized as X-stabilizers and Z-stabilizers. During the execution, measurement circuits are periodically executed to detect the errors. 
The time for executing one measurement circuit is called a surface code cycle. Surface code can be classified as double defect \cite{fowler2012surface} and lattice surgery \cite{beverland2022surface} based on the different approaches to creating logical qubits.

\subsubsection{Double Defect Model}

In double defect model, a logical qubit is created by turning off two defects of the same type. According to the type of defects, the logical qubit is initialized into \xcut~or \zcut, as shown in Fig. \ref{x-mea} and Fig. \ref{z-mea}. 
Code distance $d$ determines the number of errors that surface code can detect and correct.
All single-qubit gates can be executed in software or locally, only involving physical qubits around its two defects under the assumption in \cite{javadi2017optimized} that a steady supply of magic state qubits is at the location of the data.
We denoted these physical qubits as tiles and the rest of the qubits on the chip as channels. 

\begin{figure}[htbp]
    \centering
    \begin{subfigure}[b]{0.45\linewidth}
        \includegraphics[width=\linewidth]{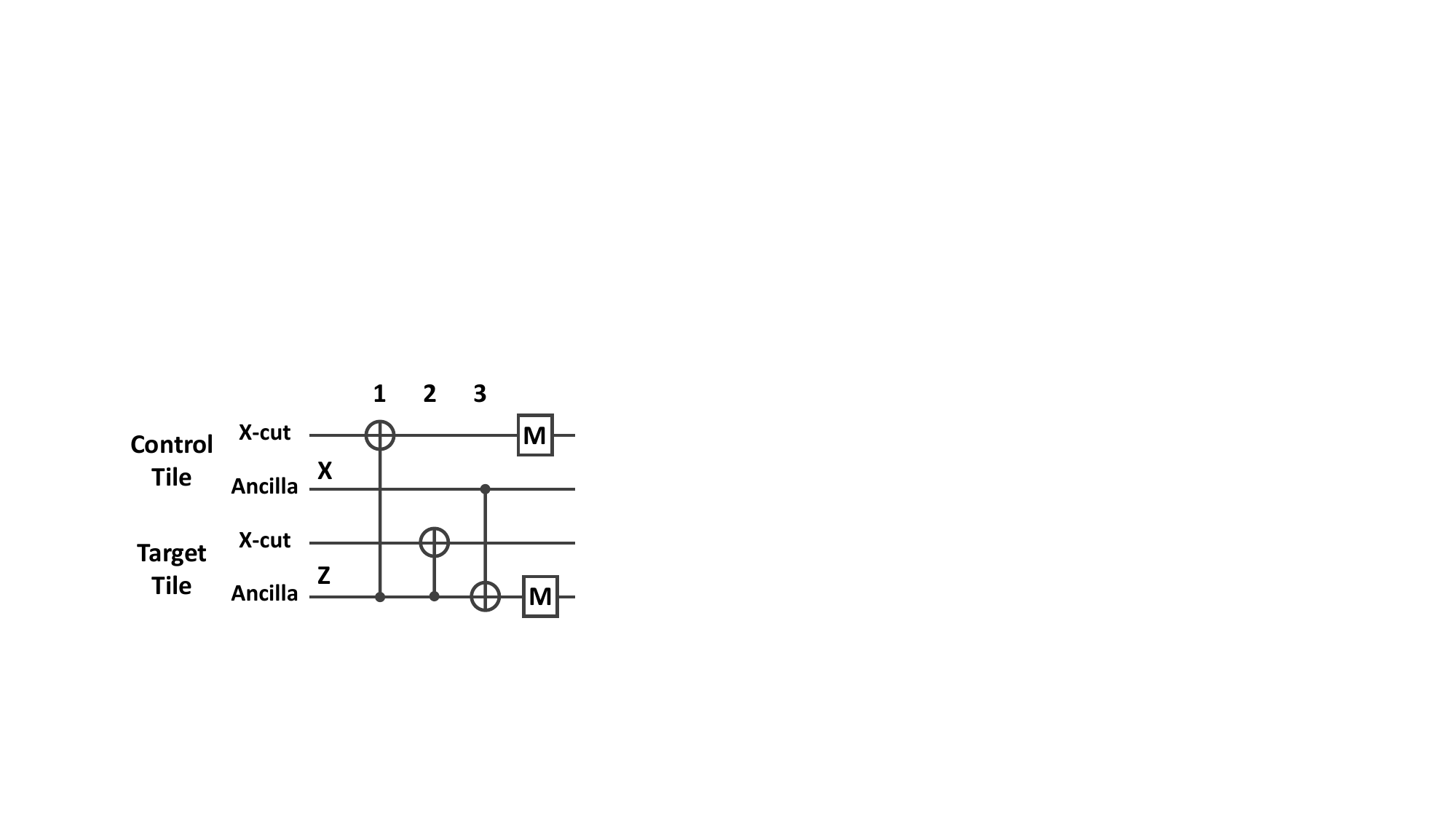}
        \caption{Direct implementation}
        \label{braiding1}
    \end{subfigure}
    \hfill
    \begin{subfigure}[b]{0.45\linewidth}
        \includegraphics[width=\linewidth]{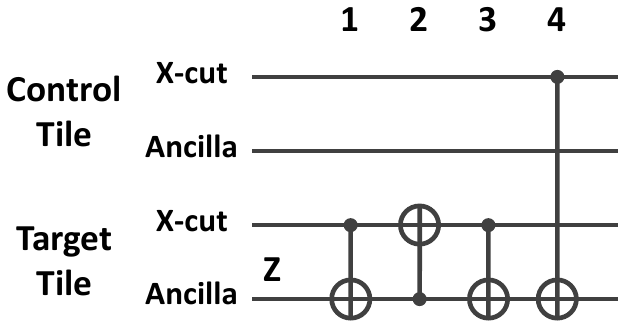}
        \caption{Modifying implementation}
        \label{braiding2}
    \end{subfigure}
    
    \caption{\cnot~gate between logical qubits of same cut type: (a) three braiding operations without cut type changing, (b) changing the cut type and executing the \cnot~gate.}
	\label{braiding}
\end{figure}

\cnot~gate requires communication between the control and target qubit, achieved by performing braiding operations in the channels. 
A braiding operation turns off the involved measurement qubits on the braiding path. It follows the topological rules: the braiding paths are equivalent as long as the starting and ending tiles are the same. Braiding operations of any length can be executed within $2d$ surface code cycles, equivalent to one clock cycle.
The braiding operation can only be performed between logical qubits with different cut types. In practice, each tile contains two double-defect logical qubits, one for computation and one for ancilla. There are two ways to perform a CNOT gate with qubits of the same cut type. One is to use three braiding operations with ancilla qubit, as shown in Fig. \ref{braiding1}. The other is to modify the cut type of the tile and then perform the braiding operation, as shown in Fig. \ref{braiding2}. They require three cycles and four cycles, respectively.

\subsubsection{Lattice Surgery Model}

\begin{figure}[htbp]
	\centering  %图片全局居中
	\includegraphics[width=0.85\linewidth]{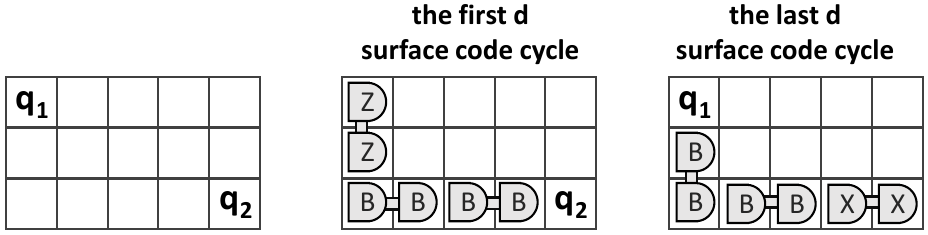}
	\caption{The CNOT gate implementation in lattice surgery model by constructing Bell states.}
	\label{lattice_surgery}
\end{figure}

\begin{figure}[htbp]
    % \centering
    \hspace{0.12\linewidth}
    \begin{subfigure}[b]{0.26\linewidth}
        \includegraphics[width=\linewidth]{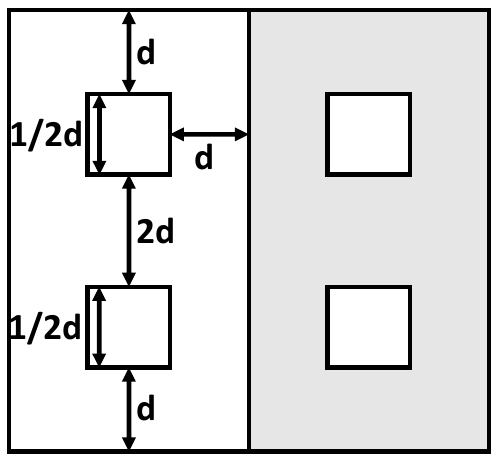}
        \subcaption{} % 添加子图标题，不计入序号
        \label{fig:dd_model}
    \end{subfigure}
    \hfill
    \begin{subfigure}[b]{0.25\linewidth}
        \includegraphics[width=\linewidth]{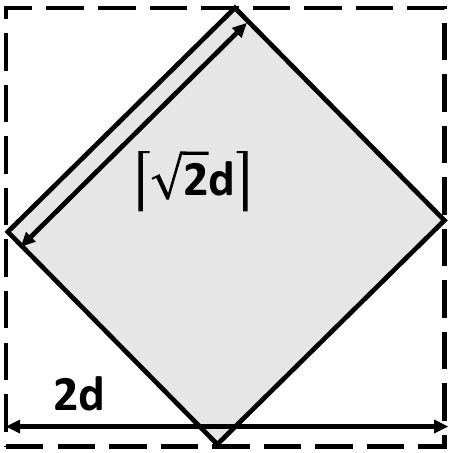}
        \subcaption{} % 添加子图标题，不计入序号
        \label{fig:ls_model}
    \end{subfigure}
    \hspace{0.12\linewidth}
    \caption{Simplified tile models: (a) double defect model, (b) lattice surgery model.}
\end{figure}

Lattice surgery eliminates the holes within tiles and uses the rotated surface code (as shown in Fig. \ref{fig:ls_model}) to reduce the requirement of qubit resources for surface code with the same code distance. CNOT gates in lattice surgery are attained by conducting ZZ measurements between neighboring tiles. A straightforward approach for a CNOT gate at a distance $k$ involves continuously swapping logical qubits until they are adjacent, requiring a minimum of $k\times d$ surface code cycles to complete. Another method involves constructing Bell states using ancilla qubits for execution, achievable within $2\times d$ surface code cycles i.e. one cycle as shown in Fig. \ref{lattice_surgery}.

%% file: 3-ProblemFormulation.tex
\section{System Model and Problem Formulation}
\label{problem_hard}
In this section, we formally define the surface code mapping and scheduling problem for both double defect and lattice surgery models and demonstrate the complexity of the problem under double defect model.

\begin{figure}[htbp]
    \centering
    \begin{subfigure}[b]{0.35\linewidth}
        \includegraphics[width=\linewidth]{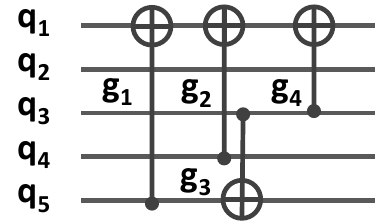}
        \subcaption{} % 添加子图标题，不计入序号
        \label{fig:Quantum circuit}
    \end{subfigure}
    \begin{subfigure}[b]{0.35\linewidth}
        \includegraphics[width=\linewidth]{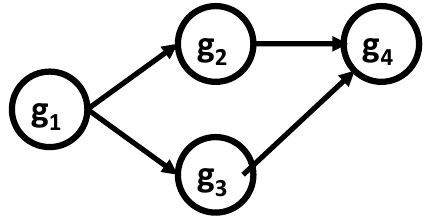}
        \subcaption{} % 添加子图标题，不计入序号
        \label{fig:DAG representation}
    \end{subfigure}
    \begin{subfigure}[b]{0.2\linewidth}
        \includegraphics[width=\linewidth]{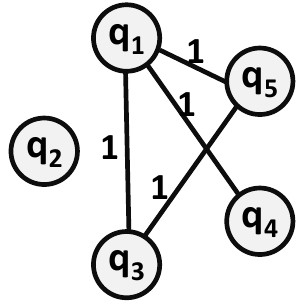}
        \subcaption{} % 添加子图标题，不计入序号
        \label{fig:Communication Graph}
    \end{subfigure}
    
    \caption{Three representations of quantum circuit: (a) original, (b) DAG, (c) communication graph.}
    % \label{fig:sidebyside}
\end{figure}

\noindent\textbf{Quantum circuit:} We consider an input quantum circuit $P$ with $n$ logical qubits (Fig. \ref{fig:Quantum circuit}). Since single-qubit gates can be implemented by software or locally in tile, we only consider CNOT gates in this work. Generally, $P$ can be represented as a directed acyclic graph (DAG) $G_P$, as shown in Fig. \ref{fig:DAG representation}. In $G_P$, each node is a CNOT gate, and edges indicate the dependency between gates. The critical path length of $G_P$ is the circuit depth, denoted as $\alpha$. The communication graph $G_C$ is another representation of a quantum circuit, as shown in Fig. \ref{fig:Communication Graph}, where each vertex is a logical qubit, and edges indicate CNOT gates between the qubits, and the weight of the edge is the number of the corresponding CNOT gates.
% how many CNOT operations there are between them.

\noindent\textbf{Quantum chip:} We assume that the topology of a quantum chip is the 2D lattice of the physical qubit, where each qubit is associated with four adjacent qubits, except the qubit on the chip's boundary. We use $L_{m_1 \times m_2}$ to denote the 2D chip with $m_1$ rows and $m_2$ columns of physical qubits.

\noindent\textbf{Surface Code Encoded Circuit:} The encoded circuits $P^S$ should satisfy the following two constraints. First, the execution scheme should be equivalent to the logical circuit, i.e., all gates are scheduled, and the scheduling order is consistent with the topological sort of gates in $G_P$. Second, the CNOT paths of the gates executed in one cycle do not intersect. The execution time of a circuit is $\Delta \times 2d \times \tau$, where $\Delta$ is the cycle number and $\tau$ is the execution time of each surface code cycle. Since $d$ and $\tau$ have the same effect on different mapping and scheduling methods, we simplify the execution time as cycle number $\Delta$.

\noindent\textbf{Surface Code mapping and scheduling Problem: }
Given an input quantum circuit $P$, a specific quantum chip $L_{m_1\times m_2}$ and the required code distance $d$ find an initial mapping and the execution scheme that satisfy the surface code circuit constraints with the cycle number of circuit $\Delta_{P^S}$ be minimized.

Next, we will refine the models for double defect and lattice surgery approaches. We further provide detailed descriptions of each model and offer formal problem definitions.

% \begin{figure}[htbp]
%     % \centering
%     \hspace{0.12\linewidth}
%     \begin{subfigure}[b]{0.26\linewidth}
%         \includegraphics[width=\linewidth]{fig/dd_model.pdf}
%         \subcaption{} % 添加子图标题，不计入序号
%         \label{fig:dd_model}
%     \end{subfigure}
%     \hfill
%     \begin{subfigure}[b]{0.25\linewidth}
%         \includegraphics[width=\linewidth]{fig/ls_model.pdf}
%         \subcaption{} % 添加子图标题，不计入序号
%         \label{fig:ls_model}
%     \end{subfigure}
%     \hspace{0.12\linewidth}
%     \caption{Simplified Tile Models: (a) double defect model, (b) lattice surgery model.}
% \end{figure}

\subsection{Double Defect Model}

\noindent\textbf{Tile:} A tile is a square array of $5d \times 5d$ physical qubits, as shown in Fig. \ref{fig:dd_model}, each tile contains two logical qubits, one for mapping logical qubits and one for \emph{ancilla}. We use $(T_{a,b},Cut_i)$ to denote tile $T_i$, where $(a,b)$ is the position of the upper left corner of the tile and $Cut_{i}$ is its cut type.

\noindent\textbf{Channel:} Channel is used to perform braiding operations. Each braiding path requires a width of $2.5d$ physical qubits. We consider the occupation of a braiding path within a channel as a lane. We introduce \textbf{bandwidth} to characterize the number of lanes in each channel. The bandwidth of a channel $C_i$ is $\lfloor \frac{W_i}{2.5d} \rfloor$, where $W_i$ is the number of physical qubits in the width of the channel $C_i$. Then, we consider the minimum bandwidth of channels within the chip as the chip's bandwidth.

\begin{figure}[htbp]
	\centering  %图片全局居中
	\includegraphics[width=0.85\linewidth]{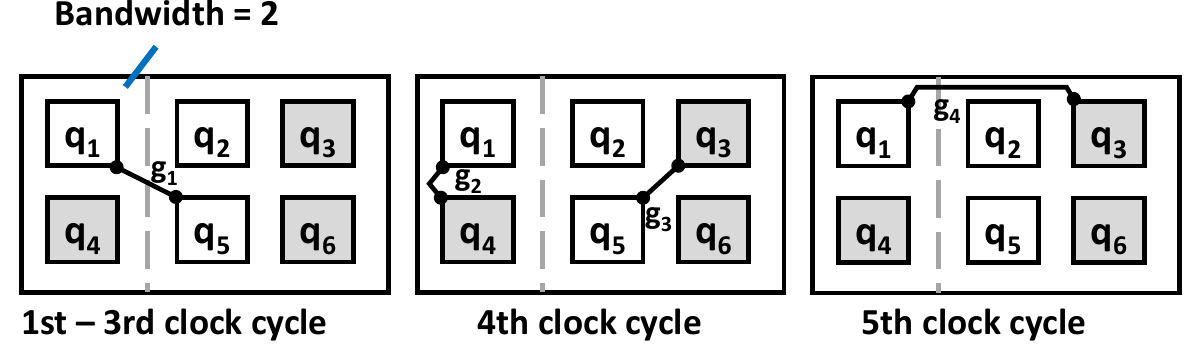}
	\caption{A five-step execution scheme for the quantum circuit in Fig.\ref{fig:Quantum circuit} using the double defect model, where the gray boxes are for X-cut tiles and the white boxes are for Z-cut tiles.}
	\label{braiding-compiler}
\end{figure}

Fig. \ref{braiding-compiler} illustrates the process of surface code transformation for the double defect model. We present the formal description as follows.
% Following, we present a formal description.

\begin{problem}
\label{problem_initial_def_dd}
\problemtitle{Initialization Problem for Double Defect.}\\
\emph{\textbf{Input:}} An input logical circuit $P$, a 2D lattice chip $L_{m_1 \times m_2}$, the required code distance $d$ and a natural number $k$. \\
\emph{\textbf{Output:}} Whether there is an initial tile mapping $T_{mapping}=\{q_{i} \rightarrow (T_{a,b},Cut_i)\}$ such that the number of cycles of the optimal surface code encoded circuit $P^{S}_{OPT}$ is upper bounded by $\alpha + k$, namely, $\Delta^{P^{S}_{OPT}}< \alpha + k$.
\end{problem}

\begin{problem}
\label{problem_Gate_def_dd}
\problemtitle{Scheduling Problem for Double Defect.}\\
\emph{\textbf{Input:}} An input logical circuit $P$, a 2D lattice chip $L_{m1 \times m2}$ and an initial tile mapping $T_{mapping}$.\\
\emph{\textbf{Output:}} A surface code encoded circuit $P^S$ with its number of cycles $\Delta^{P^{S}}$ minimized.
    
\end{problem}

\noindent\textbf{Hardness: }

\begin{theorem}
\label{hardness}
The surface code tile initialization problem for double defect model is NP-hard.
\end{theorem}

\textbf{Proof sketch:} 

We reduce the Initialization Problem for double defect model into a 3-SAT problem. For more details, please refer to Appendix \ref{appendix:A}.

\subsection{Lattice Surgery Model}

\noindent\textbf{Tile:} 
As shown in Fig. \ref{fig:ls_model}, each small box represents a tile that can be mapped as a logical qubit, with $\lceil \sqrt{2}d\rceil \times \lceil \sqrt{2}d\rceil$ physical qubits. We denote tile $T_i$ by $T_{a,b}$, where $(a,b)$ represents the upper left position of this tile.

\noindent\textbf{Channel:} Each channel is composed of tiles, which are ancilla logical qubits to generate Bell states for communication. Since both logical qubits and channels are constructed from tiles, the width of a path and a tile are the same, consisting of $d$ physical qubits. The bandwidth of the channels $C_i$ is given by $\lfloor \frac{W_i}{\lceil \sqrt{2}d\rceil} \rfloor$. The chip's bandwidth is the minimal bandwidth of its channels.

\begin{figure}[htbp]
	\centering  %图片全局居中
	\includegraphics[width=0.9\linewidth]{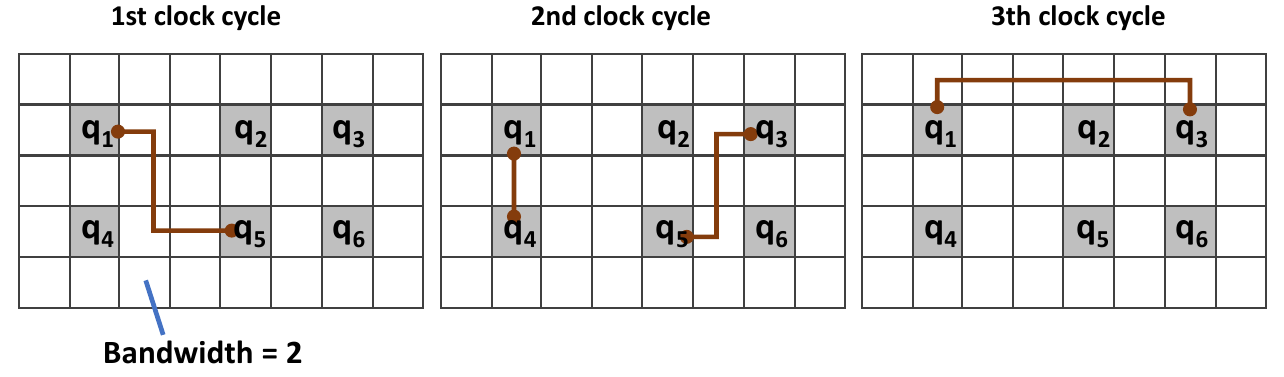}
	\caption{A three-step execution scheme for the quantum circuit in Fig. \ref{fig:Quantum circuit} using the lattice surgery model.}
	\label{ls-compiler}
\end{figure}

Fig. \ref{ls-compiler} shows the process of surface code transformation for the lattice surgery model. Below, we present the formal depiction of these problems.

\begin{problem}
\label{problem_initial_def}
\problemtitle{Initialization Problem for Lattice Surgery}\\
\emph{\textbf{Input:}} An input logical circuit $P$, a 2D lattice chip $L_{m_1 \times m_2}$, the required code distance $d$ and a natural number $k$. \\
\emph{\textbf{Output:}} Whether there is an initial tile mapping $T_{mapping}=\{q_{i} \rightarrow T_{a,b}\}$ such that 
$\Delta^{P^{S}_{OPT}}< \alpha + k $, namely, the number of cycles of the optimal surface code encoded circuit $P^{S}_{OPT}$ is upper bounded by $\alpha + k$.
\end{problem}

\begin{problem}
\label{problem_Gate_def}
\problemtitle{Scheduling Problem for Lattice Surgery.}\\
\emph{\textbf{Input:}} An input logical circuit $P$, a 2D lattice chip $L_{m_1 \times m_2}$ and an initial tile mapping $T_{mapping}$.\\
\emph{\textbf{Output:}} A surface code encoded circuit $P^S$ with its number of cycles $\Delta^{P^{S}}$ minimized.
    
\end{problem}

\noindent\textbf{Hardness: } Herr \etal \cite{herr2017optimization} demonstrated that the complexity of surface code mapping and transforming problem is NP-complete for lattice surgery model.

%% file: 5-SysDesign.tex
\section{System Design}
\label{sec:system}
\begin{figure}[h]
	\centering  %图片全局居中
	\includegraphics[width=\linewidth]{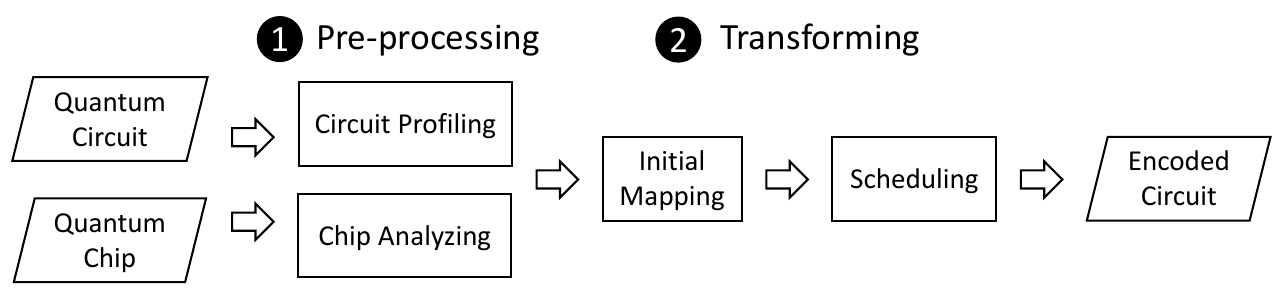}
	\caption{Overview of \ourframework.}
	\label{framework}
	\vspace{-0.4cm}
\end{figure}

It is a non-trivial task to optimize circuit mapping and scheduling on limited qubit resources.
To address the problem, firstly, we introduce two novel metrics: \emph{\para}~and \emph{\chipcapacity}~(Section. \ref{sec:Pre-processing}) to characterize quantum circuits and chips. Then, we propose resource-adaptive algorithms  \ourframework~(Section. \ref{sec:compiling}) with customized initialization of chip resources for each circuit. Further, with sufficient physical qubits on the chip, \ourframework-\unlimit~ can have a shorter transforming time and performance-guaranteed result. 
An overview of our comprehensive toolflow is shown in Fig. \ref{framework}.

\subsection{Pre-processing}
\label{sec:Pre-processing}
\subsubsection{Quantum Circuit Profiling}
\label{Quantum-Circuit-Profiling}
Different quantum circuits may have various demands on communication resources. We introduce \emph{\para}~(denoted as $\mathbb{PM}$) to characterize the maximum demand of communication resources of a circuit.

\begin{definition}
\emph{\para}: Given a quantum circuit $P$, $G_P=(V,E)$. A partition $\pi$ is to divide nodes $v \in V$ into $\Delta_{G_P}$ disjoint set $V_1,V_2,...,V_{\Delta_{G_P}}$, such that for $u \in V_i$ and $ v \in V_j$, if $(u,v) \in E $, then $i > j$. $\mathbb{PM}=\min_{\pi}\max_{i=1}^{\Delta_{G_P}}|V_i|$
\end{definition}

Finding $\mathbb{PM}$ is equivalent to given $n$ tasks and their precedence constraints, minimizing the number of machines used while the whole schedule is of minimum length. Finke \cite{finke2009minimizing} has proved that this is NP-complete.

We propose a heuristic algorithm (Algorithm Para-Finding) to find the circuit's estimate \para~$\widetilde{\mathbb{PM}}$ and the corresponding execution order. Our methods use layers to keep track of the execution order of gates, where $layer_1$ represents the operations to be performed in the first time cycle. For any gate $i$, we record two values, the highest and lowest layers, that the gate can be scheduled. Layers are determined by the gate's parents and children nodes, denoted as $parent_i$ and $child_i$. $Low_{i}=\max_{j\in parent_i}Low_j+1$ and $High_{i}=\min_{j\in child_i}High_j-1$.
Then, we calculate the difference between the gates' high and low values and choose the gate with the smallest difference. For this gate, we schedule it to the layer with the fewest gates to execute in all the possible layers.
After that, we update the low value of its child nodes and the high value of its parent node. We repeat this process until all the gates are scheduled. The maximum number of gates per layer is $\widetilde{\mathbb{PM}}$ of this circuit.

\subsubsection{Quantum Chip Analyzing}

\label{chip_theo}

We define the \emph{\chipcapacity}~to characterize the number of parallel CNOT gates supported by a chip, denoted as $\mathbb{C}$. According to \cite{hua2021autobraid}, any three CNOT gates can be executed simultaneously. As we refine the chip model, we generalize the previous theorem to the case that any $\lfloor \frac{b-1}{2}\rfloor +3$ braiding operations can be executed simultaneously where $b$ is the chip's bandwidth.

\begin{definition}
\emph{\chipcapacity: }Given a quantum chip, $\mathbb{C}$ is the max number $u$ that for any $u$ independent \cnot~gates with an arbitrary placement of tiles, there exists a simultaneous path schedule for all \cnot~gates.
\end{definition}

\begin{theorem}
\label{chip}
For a chip with bandwidth $b$, given an arbitrary placement of the operand qubits, there exists a simultaneous paths schedule for $\lfloor \frac{b-1}{2}\rfloor +3$ gates.
\end{theorem}

\noindent \textbf{Proof:} 
According to Autobraid\cite{hua2021autobraid}, any three CNOT operations must be able to execute simultaneously on a chip with bandwidth 1. The path of the additional CNOT gate has to intersect with others if and only if one involved tile is inside a ring and the other is outside. A ring is composed of paths and fully occupied channels. Increasing the bandwidth of each channel on the chip by two would break this ring and enable a path connecting two arbitrary tiles on the chip. Therefore, when the chip's bandwidth is $b$, paths exist for $\lfloor \frac{b-1}{2}\rfloor +3$ CNOT operations to be executed in parallel.

\subsection{Transforming}
\label{sec:compiling}
\subsubsection{Initial Mapping}
\begin{figure}[htbp]
    \centering
    
    \begin{subfigure}[b]{0.48\linewidth}
        \includegraphics[width=\linewidth]{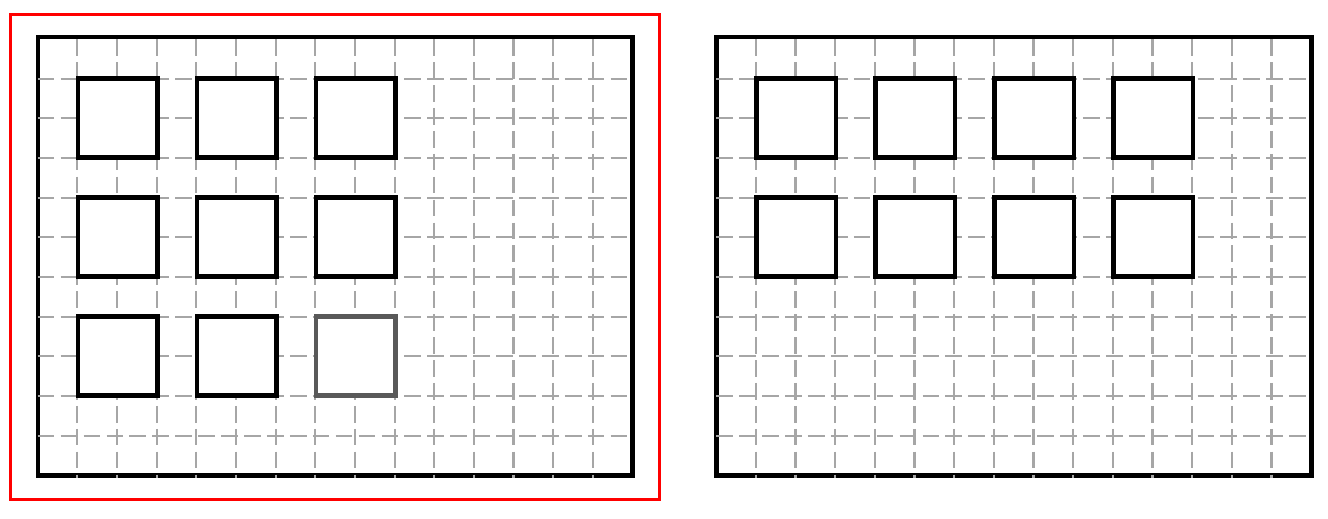}
        \subcaption{} % 添加子图标题，不计入序号
        \label{location1}
    \end{subfigure}
    \hfill
    \begin{subfigure}[b]{0.23\linewidth}
        \includegraphics[width=\linewidth]{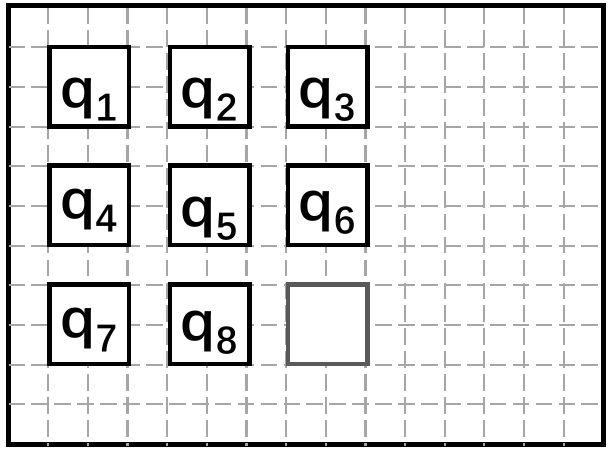}
        \subcaption{} % 添加子图标题，不计入序号
        \label{location2}
    \end{subfigure}
    \hfill
    \begin{subfigure}[b]{0.23\linewidth}
        \includegraphics[width=\linewidth]{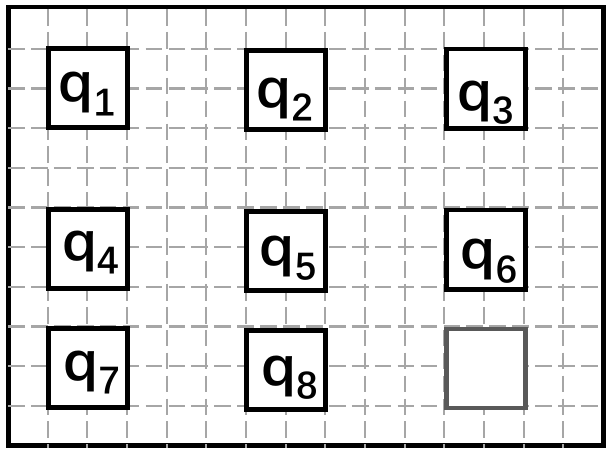}
        \subcaption{} % 添加子图标题，不计入序号
        \label{location3}
    \end{subfigure}
    
    \caption{Tile location mapping process: (a) Shape determining, (b) mapping establishing, (c) bandwidth adjusting.}
    \label{fig:location_process}
\end{figure}

To generate a preferred tile location mapping, we employ the following three steps (Line\ref{line:mapping1}-\ref{line:mapping3} in Algorithm\ref{alg:InSufficient}): 
% 1) determining the shape of physical qubits for encoding logical qubits; 2) establishing a mapping between logical qubits and corresponding tiles; 3) adjusting the bandwidth of channels.

\noindent \textbf{Shape Determining.} First, we determine the shape of the logical tile array, i.e., whether to initialize it as a $3\times3$ array or a $2\times4$ array for a circuit with eight logical qubits when both schemes are available on the chip. Then, We select an array shape with the minimum perimeter. As shown in Fig. \ref{location1}, we choose the $3\times3$ tile array.

\noindent \textbf{Mapping Establishing.} Secondly, We map each logical qubit to its corresponding logical tile according to communication cost, as shown in Fig. \ref{location2}. The communication cost is calculated by cost function $f=\sum_{i,j} \gamma_{i,j} \times l_{i,j}$, where $l_{i,j}$ represents the Manhattan distance between the two tiles $T_i$ and $T_j$, and $\gamma_{i,j}$ is the number of $CNOT_{i,j}$ in the circuit. Mapping qubits that frequently communicate together can effectively reduce the communication cost. Here, we employ the Metis\cite{karypis1997metis} method, an iterative graph partitioner, to generate mappings based on the qubit communication graph $G_C$ and tile array. Due to the stochastic steps in the mapping generation, we generate multiple mappings and select the one with minimal communication cost as our final result.

\noindent \textbf{Bandwidth Adjusting.} 
Finally, we assign the rest of the qubit resources to each channel based on their occupancy status, as illustrated in Fig. \ref{location3}.
We pre-execute each gate in the circuit to record its shortest path without considering the non-intersecting restrictions. Then, we increase the bandwidth for channels that perform the most paths. In most cases, this process effectively reduces the wait caused by channel resource occupation.

\begin{algorithm}
    \caption{Scheduling for Limited Resources} \label{alg:InSufficient}
    \KwIn{Quantum Circuit $P$ and Chip $L_{m_1\times m_2}$}
    \KwOut{Encoded Circuit $P^S$}
    tile\_array = Tile\_shaping($L_{m_1\times m_2}$);\label{line:mapping1}\\
    Mappings = Metis($G_C$);\\
    M\_location = Select(Mappings,cost function);\label{line:mapping3}\\
    \If{double defect model}{
        M\_cut = bipartite($G_P$);
    }
    % % Initialize $\mathbb{S}$ and chip\_now;\\
    \While{$G_P$ not empty}{ \label{line:sch}
        gates=$G_P$.front\_gate\\
        gates\_pri = priority(gates);\\
        \For{$g_i(q_a,q_b) \gets gates\_pri.begin()$ \KwTo $gates\_pri.end()$}{
            \eIf{$Cut_a \neq Cut_b$ or model = lattice surgery}{
            $P^S$.add(path(gate, chip\_now));\label{line:sch2}
            }
            {
            $M_a = {M_t}_a+\theta {M_s}_a$;\\ \label{line:cut_limit1}
            $M_b = {M_t}_b+\theta {M_s}_b$;\\
            $min\_value,min\_index = \min(M_a,M_b)$;\\
            \eIf{$min\_value<0$}
            {$P^S$.add($Cut_{min\_index}$ modification);}
            {$P^S$.add(path(gate, chip\_now));}
            }
            }\label{line:cut_limit2}
        }
    return $P^S$, M\_location
\end{algorithm}

\subsubsection{Scheduling}
\label{alg:scheduling}
%针对资源是否充足的两种情况，我们设计了两种算法来最大化利用资源。
Considering the qubit resources on the target chip may be limited or sufficient, we design two algorithms to maximize the utilization of resources.

\noindent \textbf{Scheduling for limited Resources.}
When the resources of physical qubits are limited, i.e., when $\widetilde{\mathbb{PM}} > \lfloor \frac{b-1}{2}\rfloor +3$, it may be difficult to find non-intersecting paths to execute all current executable gates.
However, the number of children in gates of currently executable gates varies. We assign priorities to these nodes, effectively reducing latency at the bottleneck (Line \ref{line:sch} - \ref{line:sch2} in Algorithm \ref{alg:InSufficient}).
The priority of a gate is determined by the remaining gates number (how many gates depend on it) and criticality (the length of the critical path of the remaining gates). Gates with higher criticality are prioritized. When two \cnot~gates have the same criticality, we select the gate with more remaining gates to allow more gates to execute earlier to utilize non-congested cycles better. 

The time complexity of this algorithm is $O(g * m_1 * m_2)$, where $g$ is the number of CNOT gates in the quantum circuit. The algorithm searches for paths for at most $O(g)$ gates, and the maximum time required to find a path for each CNOT gate is $O(m_1 * m_2)$. Here, $O(m_1 * m_2)$ is the number of nodes available for path selection on the chip, which is $m_1 * m_2 / d * d$,  where $d$ is the code distance and the side length of each tile.

\begin{algorithm}
\caption{Scheduling for Sufficient Resources} \label{alg:Sufficient-alg}
\KwIn{Execution Scheme $E$}
\KwOut{Encoded Circuit $P^S$, initialization}
% $\text{init}\quad M_{l},G$ and $\mathbb{S}$ 
%\tcp{They are trivial location mapping, empty graph, and result records respectively}
now\_step = 0;\\
\While{$i < E.length()$}{
    \While{$G$ is bipartite graph}{\label{line:cut1}
    \For{$gate \gets E[i].begin()$ \KwTo $E[i].end()$}{
        G.add\_edge(gate);
        }
    i++;\\
    $M_{c}$ = bipartite($G$);
    }\label{line:cut2}
    \eIf{having mapping}{$P^S$.add(change mapping to $M_{c}$);}{initialization = $M_{c}$}
    \For{$j \gets now\_step$ \KwTo $i$}{
        find braiding path($E[j]$);\\
        $P^S$.add($E[j]$);
        }
    now\_step = i;
}
return $P^S$, initialization
\end{algorithm}

\noindent \textbf{Scheduling for Sufficient Resources.}
When $\lfloor \frac{b-1}{2}\rfloor +3 \geq \widetilde{\mathbb{PM}}$, an execution scheme can be rapidly derived from Algorithm Para-Finding and Theorem \ref{chip}. Algorithm Para-Finding provides $\widetilde{\mathbb{PM}}$ for this quantum circuit and a CNOT gate order scheme that achieves $\widetilde{\mathbb{PM}}$. This scheme indicates which gates are executed in each time cycle. Since the number of gates executed in each cycle is smaller than $\lfloor \frac{b-1}{2}\rfloor +3$, employing the methods outlined in Proof \ref{chip} to determine the corresponding paths for these gates becomes feasible. 
% For each gate, denoted as $g_1$, $g_2$, ..., $g_k$, paths are sought in the $1^{st}$ and $(2\times k -1)^{th}$ lane of all channels, the $2^{nd}$ and $(2\times k -2)^{th}$ lanes, and subsequently, the $k^{th}$  lanes.

\subsection{Optimizations for Double Defect Model}
%针对double defect model,我们提出了cut type初始化和调度方法，用于减少每个cnot门的执行时间。
Previous works Braidflash \cite{javadi2017optimized} and Autobraid \cite{hua2021autobraid} do not consider the cut type by assuming all tiles have the same cut type. However, cut type is critical in transforming for double defect model, providing a significant opportunity to reduce the time on the table. For a \cnot~gate, it takes \textbf{three} cycles to be executed if two involved tiles are of the same cut type, but only \textbf{one} cycle when cut types are different.

\subsubsection{Cut Type Initialization}
\ 
\newline
\indent The goal of the cut type initialization is to enable the execution of as many CNOT gates as possible within a single cycle. If the qubit communication graph is bipartite, we assign the same cut type to the logical qubits in the same set. 
This is the optimal cut type initialization, with which all \cnot~gates can be implemented in one cycle. 

However, for circuits whose qubit communication graph is not bipartite, find the optimal cut type initialization is NP-hard, according to Theorem \ref{hardness}. We propose a greedy algorithm that satisfies the requirement of cut type for gate executed earlier.
%since the gate executed later can change the tile's cut type. 
To end this, firstly, we construct a sub-graph of the qubit communication graph where each vertex corresponds to a logical qubit. Then, we add the gates with no precursor in the current dag into the sub-graph. Next, we remove these gates in the DAG. Repeat these two steps until the newly added edges make the sub-graph no longer bipartite. The logical qubits belonging to the same set in this bipartite sub-graph are initialized to the same cut type. 

\subsubsection{Scheduling}
When involving two tiles of a CNOT gate are of the same cut type, we estimate the impact of modifying cut type by calculating the M-value of each tile, specifically $ \text{M-value} = M_{t} + \theta \times M_{s}$. $M_{t}$ is the impact on time. It takes three cycles to execute the operation directly and four cycles with modification. If this tile is idle previously, the modification operation can be performed earlier to reduce the time cost. $M_{s}$ is the impact of the occupation of the channel. \cnot~gate needs two braiding operations between the tiles without changing the cut type but only needs one after modification. We adopt the look-forward strategy considering the impact of this modification on the children gates of this gate.
% Note that we only consider children nodes that may be executed within three braiding cycles. Otherwise, tiles can easily modify their cut type in the interval of execution. 
The parameter $\theta$ is used to determine the weights of the two factors, $M_{t}$ and $M_{s}$, in the current situation, where $\theta = (|\text{ready gate}|\times 2)/\text{bandwidth} \times n$.
We choose to modify the type of the tile when the M-value is greater than 0 (Line \ref{line:cut_limit1} - \ref{line:cut_limit2} in Algorithm \ref{alg:InSufficient}).

\subsubsection{Sufficient Scheduling}
When physical qubit resources are sufficient, we adopt the methods in Section \ref{alg:scheduling} to determine the tile location mapping and gate schedule scheme. The key idea for cut type initialization and scheduling is to make all \cnot~gates execute in one cycle by remapping the cut type. 

We propose the cut type scheduling algorithm Algorithm \ref{alg:Sufficient-alg}, whose execution flow is as follows.
Firstly, we construct the qubit communication graph by sequentially adding edges from the execution scheme until it is no longer bipartite. Then, we use this bi-partition graph to initialize the cut type for executing this sub-circuit. When the operand tiles of \cnot~gate are of the same cut type, our methods spend three cycles to modify the cut type to the new mapping found in the same way above. These two steps are iterated over until all the gates have been scheduled. 
We provide the cut type scheduling algorithm with $\frac{5}{2}$-approximation guarantee (as shown in Theorem \ref{theo:5/2}).
\begin{lemma}
\label{bipartite_graph}
The qubit communication graph generated by any two layers of gates is bipartite.
\end{lemma}

\noindent \textbf{Proof:} Since logical qubits can participate in at most one \cnot~gate in each layer, the qubit communication sub-graph generated by the 2-layer circuit has a maximum degree of two, and the two edges connected by a vertex must belong to two different layers. A graph with a maximum degree of two can only consist of lines or rings. A ring must be an even ring since two edges must be connected by a vertex in the odd ring that belongs to the same layer. As a result, this graph is bipartite since the qubit communication sub-graph can only consist of lines and even rings.

\begin{theorem}
\label{theo:5/2}
Algorithm \ref{alg:Sufficient-alg} is $\frac{5}{2}$-approximation.
\end{theorem}

\noindent \textbf{Proof:} For every two cycles of gates in the execution scheme given by Algorithm Para-Finding, the optimal cases must take two braiding cycles to execute since the gates with gate dependencies cannot be executed simultaneously. According to Lemma \ref{bipartite_graph}, our method requires at most five braiding cycles to execute these two layers of gates, three cycles for modifying to optimal cut type mapping, and two cycles for performing braiding operations. Thus, our algorithm is $\frac{5}{2}$-approximation.

%% file: 6-Evaluation.tex
\section{Performance Evaluation}
\label{sec:experiment}

\begin{table*}[ht]
\centering
\renewcommand\arraystretch{1}
\caption{Overview of Experiment Results} 
\label{table: main results}
\begin{threeparttable}
\resizebox{0.85\linewidth}{!}{

\begin{tabular}{c c c c  c c c  c c c c}
\toprule
\multirow{2}*{Circuit} & \multirow{2}*{$n$} & \multirow{2}*{$\alpha$} &\multirow{2}*{$g$\footnotemark[1]} & \multicolumn{1}{c}{Autobraid} &
\multicolumn{2}{c}{\ourframework-dd}&
\multicolumn{2}{c}{\PRX}&\multicolumn{2}{c}{\ourframework-ls} \\

& & & & Min & Min  &  \unlimit & Min & 4X & Min &  4X \\

\midrule
dnn\_n8 & 8 & 48 & 192 & 147 & \textbf{48} & 48 & 48 & 53 & \textbf{48} & \textbf{48} \\
grover & 9 & 110 & 132 & 330 & \textbf{166} & 140 & 110 & 110 & \textbf{110} & \textbf{110} \\
qpe\_n9 & 9 & 42 & 43 & 126 & \textbf{70} & 54 & 42 & 42 & \textbf{42} & \textbf{42} \\
BV\_10 & 10 & 5 & 5 & 15 & \textbf{5} & 5 & 5 & 5 & \textbf{5} & \textbf{5} \\
QFT\_10 & 10 & 93 & 105 & 279 & \textbf{165} & 96 & 93 & 93 & \textbf{93} & \textbf{93} \\
adder\_n10 & 10 & 55 & 65 & 165 & \textbf{78} & 82 & 55 & 56 & \textbf{55} & \textbf{55} \\
ising\_n10 & 10 & 20 & 90 & 60 & \textbf{20} & 20 & 20 & 20 & 24 & \textbf{20} \\
sat\_n11 & 11 & 204 & 252 & 612 & \textbf{336} & 339 & 204 & 204 & \textbf{204} & \textbf{204} \\
square\_root\_n4 & 11 & 221 & 294 & 663 & \textbf{379} & 389 & 221 & 225 & \textbf{221} & \textbf{221} \\
multiplier\_n15 & 15 & 133 & 222 & 399 & \textbf{232} & 244 & 133 & 134 & \textbf{133} & \textbf{133} \\
qf21\_n15 & 15 & 112 & 115 & 336 & \textbf{197} & 130 & 112 & 112 & \textbf{112} & \textbf{112} \\
dnn\_n16 & 16 & 48 & 384 & 198 & \textbf{71} & 48 & 79 & 53 & \textbf{68} & \textbf{52} \\
square\_root\_n18 & 18 & 644 & 898 & 1932 & \textbf{1047} & 1133 & 644 & 645 & \textbf{644} & \textbf{644} \\
ghz\_state\_n23 & 23 & 22 & 22 & 66 & \textbf{22} & 22 & 22 & 22 & \textbf{22} & \textbf{22} \\
multiplier\_n25 & 25 & 381 & 670 & 1143 & \textbf{659} & 717 & 383 & 385 & \textbf{381} & \textbf{381} \\
swap\_test\_n25 & 25 & 63 & 96 & 201 & \textbf{89} & 99 & 67 & 65 & \textbf{63} & \textbf{63} \\
wstate\_n27 & 27 & 28 & 52 & 84 & \textbf{28} & 28 & 28 & 28 & \textbf{28} & \textbf{28} \\
BV\_50 & 50 & 27 & 27 & 81 & \textbf{27} & 27 & 27 & 27 & \textbf{27} & \textbf{27} \\
QFT\_50 & 50 & 2363 & 2435 & 7089 & \textbf{4633} & 2366 & 2363 & 2363 & \textbf{2363} & \textbf{2363} \\
ising\_n50 & 50 & 4 & 98 & 15 & \textbf{10} & 4 & 6 & 6 & 9 & 7 \\
quantum\_walk & 11 & 14104 & 14372 & 42312 & \textbf{20188} & 19669 & 14104 & 14104 & \textbf{14104} & \textbf{14104} \\
shor & 12 & 13412 & 13838 & 40248 & \textbf{22978} & 20315 & 13412 & 13414 & 13414 & \textbf{13412} \\

\bottomrule
\end{tabular}

}

\footnotesize $^1$ $n$ is the number of the qubits, $\alpha$ is the depth of the circuit, $g$ is the CNOT gate of the circuit.

\end{threeparttable}
% \vspace{-0.2cm}
\end{table*}

In this section, we first compare the performance of our methods to several state-of-the-art methods, AutoBraid \cite{hua2021autobraid} for double defect model and \PRX~\cite{beverland2022surface} for lattice surgery model. Then we evaluate the performance of \ourframework~as the communication resources increased. The details of our evaluation results are shown in Section \ref{main-result} and we highlight our key findings as follows:

\begin{itemize}
    \item For double defect model, \ourframework~outperforms Autobraid\cite{hua2021autobraid}, reducing the cycle of the transformed circuit by 67.3\% at most, on average 51.5\%.
    \item For lattice surgery model, \ourframework~reaches the optimal solution in most of the test benchmarks, reducing the cycle of the transformed circuit by 13.9\% at most compared with \PRX~\cite{beverland2022surface}.
    % \item \ourframework~yields more optimization for circuits whose parallelism is not too high, up to 40\% for both models, compared with Autobraid\cite{hua2021autobraid} and \PRX~\cite{beverland2022surface}.
    \item Compared with the result in the minimum viable chip, when the chip size increases 4x, \ourframework~reduces the execution time by 10.8\% and 30.9\% in the double defect model and lattice surgery model respectively.
    \item \ourframework~exhibits excellent scalability, effectively reducing the execution time of circuits as the chip size increases, while maintaining linear growth in compilation time.
    
    % effectively leverages the physical qubit resources on the chip, allowing for a 4x increase in physical qubits. This enhancement results in an average reduction of cycles by 10.8\% in the double defect model and 30.9\% in the lattice surgery model
\end{itemize}

\subsection{Evaluation Setting}
% depth

\noindent\textbf{Metrics.} We use the number of cycles to represent the communication time, which is used to measure the effectiveness of the compilation results. 

\noindent\textbf{Baselines.}
For double defect and lattice surgery models, we select the state-of-the-art algorithms AutoBraid\cite{hua2021autobraid} and \PRX~\cite{beverland2022surface} as our baselines.

\noindent\textbf{Chip Configuration.} We evaluate our mapping and scheduling algorithm on three resource configurations $L_{l \times l}$: minimum viable, 4x, and sufficient qubits. For minimum viable qubits, $l=\lceil\sqrt{n}\rceil \times 5d$ for double defect model and $l=\lceil\sqrt{n}\rceil \times \lceil \sqrt{2}d\rceil$ for lattice surgery model, which is the smallest square grid chip that provides enough qubits. The 4x resource for lattice surgery is a chip with $l=\lceil\sqrt{n}\rceil \times 5d$. And the $l$ for \ourframework-\unlimit~on sufficient resources depends on $\mathbb{PM}$ of the circuit.

\noindent\textbf{Benchmarks.} We use the quantum circuit from the previous works, including IBM Qiskit \cite{Qiskit}, ScaffCC \cite{javadiabhari2014scaffcc}, QUEKO \cite{tan2020optimality},  QASMbench \cite{li2022qasmbench} and random circuits with certain parallelism degree. 

\noindent\textbf{Evaluation Platform.} Our experiments are performed on Intel(R) Xeon(R) CPU 6248R 96vCores 3.00GHz, with 256GB DDR4 memory. The operating system is Ubuntu 20.04.

\subsection{Experiment Results}
\label{main-result}

\subsubsection{Double Defect Model}
We evaluate the performance of \ourframework~with two chip configurations: 1) minimum viable chip, and 2) sufficient resources. As shown in Table \ref{table: main results}, \ourframework~outperforms AutoBraid methods with a reduction by 67.3\% at most in the number of cycles, on average 51.5\%. In \ourframework-\unlimit, 40.9\% of the circuits are further reduced in the number of cycles. Compared with AutoBraid, the average cycles are reduced by 57.1\%.
Increasing communication resources addresses the latency caused by braiding congestion. Thus, only the circuits suffering braiding congestion can benefit from the increase in bandwidth. The greater the parallelism, the more the time decreases as the bandwidth increases.
\ourframework-\unlimit~is not always the best among these results. The reason is that the \ourframework-\unlimit~schedules gates have more strict limits for performance guarantee.

\subsubsection{Lattice Surgery}
For most circuits, our approach obtains the optimal solution as well as \PRX~in Table \ref{table: main results}. For circuits with higher $\mathbb{PM}$, \ourframework~achieving better results reduction cycle of up to 13.9\% compared with \PRX. 
Since the \ourframework-\unlimit~for the lattice surgery model is guaranteed to yield the optimal solution (as described in Section \ref{alg:scheduling} ), we did not evaluate its performance here.
Due to the absence of specialized optimizations in our approach for circuits with specific patterns, it falls \PRX~on the circuit ising\_n10 on the minimum viable chip, whose CNOT gates are adjacent in the snake mapping. However, the absence of an effective initial mapping hampers \PRX~capacity to capitalize on the increased physical qubit resources. Our approach can leverage additional chip resources to reduce circuit depth. All results on the 4x resources are superior to or equal to the minimal viable chip.

\subsection{Sensitivity Study}
%量子线路
In this section, we conduct sensitivity studies to investigate the impact of our strategies, i.e., location and cut type initialization, gate prioritizing, and cut type scheduling. We also analyze the scalability of circuit parallelism and chip size. These results demonstrate that our method outperforms baselines on most quantum application circuits, especially those with medium to high parallelism. Moreover, our algorithm effectively utilizes the redundant physical qubit resources on the chip to reduce the circuit cycle.

\subsubsection{Initialization Method}
We examine the impact of initialization on the execution time in location and cut type. These evaluations and the subsequent scheduling experiments are conducted on the minimum viable chip.

\noindent\textbf{Location:} As illustrated in Table \ref{table: location}, our method consistently shows superior performance in most circuits. We compared the influence of tile location selection on circuit depth with the trivial mapping in \PRX~\cite{beverland2022surface} and the Metis mapping\cite{karypis1997metis}. The trivial method refers to a twisting layout of logical qubits, where the qubits in the first row are placed from left to right, followed by the qubits in the second row placed from right to left, and repeated until all logical qubits are fully mapped. 
The shortcomings on the Ising circuit primarily result from the absence of specialized optimization targeting specific patterns. Nevertheless, the overall performance trend underscores our approach's robustness across diverse scenarios.

\begin{table}[ht]
\centering
\renewcommand\arraystretch{1}
\caption{Comparision of location initialization methods} \label{table: location}
\begin{threeparttable}
\resizebox{\linewidth}{!}{
\begin{tabular}{c c c c c c c}
\toprule
Circuit name & $n$ & $\alpha$ & $g$ & Trivial & Metis & Ours \\
\midrule
dnn\_n8 & 8 & 192 & 48 & 48 & 72 & \textbf{48} \\
grover & 9 & 132 & 110 & 112 & 110 & \textbf{110} \\
qpe\_n9  &  9  &  42  &  43  &  42  &  42  &  \textbf{42} \\
ising\_n10 & 10 & 90 & 20 & 20 & 36 & \textbf{20}\\
adder\_n10 & 10 & 65 & 55 & 55 & 55 & \textbf{55} \\
QFT\_10 & 10 & 105 & 93 & 95 & 93 & \textbf{93} \\
% shor\_a11 & 12 & 9248 & 6922 & 6924 & 6928 & \textbf{6922} \\
multiply\_n13 & 13 & 40 & 23 & 25 & 24 & \textbf{23} \\
% rd32-v0\_66 & 16 & 12 & 12 & 12 & 12 & 12 \\
square\_root\_n18 & 18 & 898 & 644 & 644 & 644 & \textbf{644} \\
ghz\_state\_n23  &  23  &  22  &  22  &  22  &  22  & \textbf{ 22} \\
swap\_test\_n25  &  25  &  63  &  96  &  63  &  63  &  \textbf{63} \\
ising\_n50 & 50 & 98 & 4 & \textbf{4} & 11 & 9 \\
% BV\_50 & 50 & 27 & 27 & 27 & 27 & \textbf{27} \\
% QFT\_50 & 50 & 2435 & 2363 & 2364 & 2363 & \textbf{2363} \\
% BV\_100 & 100 & 50 & 50 & 50 & 50 & \textbf{50} \\
\bottomrule
\end{tabular}
}

\end{threeparttable}
\end{table}

\noindent\textbf{Cut type:} In most cases, our method outperforms the baseline methods, as shown in Table \ref{table: cut type}. 
We compare our cut type initialization algorithm with the random and max-cut algorithms. The random method assigns tiles with a random cut type. The max-cut method maximizes the number of \cnot~gates with different cut types. We use the one\_exchange method in NetworkX to implement the max-cut partition. For specific circuits such as ghz\_state\_n23, our initialization algorithm can significantly reduce the number of cycles. This is because the max-cut method aims to reduce the overall number of \cnot~gates with different cut types. However, the cut type of tiles is dynamic since it can be modified during the execution. The initialization method should emphasize the front part of the quantum circuit.

\begin{table}[ht]
\centering
\renewcommand\arraystretch{1}
\caption{Comparision of cut type initialization methods} \label{table: cut type}
\begin{threeparttable}
\resizebox{\linewidth}{!}{
\begin{tabular}{c c c c c c c}
\toprule
Circuit name & $n$ & $\alpha$ & $g$ & Random & Max-cut& Ours \\
\midrule
% 4gt11\_83& 5 & 14 & 14 & 21 & 28 & \textbf{18} \\
% 4gt5\_75 & 5 & 33 & 38 & 72 & 82 & \textbf{70} \\
% alu-v0\_26 & 5 & 35 & 38 & 75 & \textbf{68} & 70\\
dnn\_n8 & 8 & 48 & 192 & 64 & 48 & \textbf{48}\\
% rd53\_251 & 8 & 492 & 564 & 1.11k & 1.06k & \textbf{1.06k} \\
grover &  9  &  110  &  132  &  173  &  172  &  \textbf{166} \\
qpe\_n9 & 9 & 42 & 43 & 73 & 76 & \textbf{70} \\
ising\_n10  &  10  &  20  &  90  &  37  &  29  &  \textbf{20} \\
adder\_n10 & 10 & 55 & 65 & 85 & 82 & \textbf{78}\\
QFT\_10  &  10  &  93  &  105  &  171  &  173  &  \textbf{165} \\
multiply\_n13 &  13  &  23  &  40  &  39  &  37  &  \textbf{35} \\
square\_root\_n18  &  18  &  644  &  898  &  1052  &  1053  &  \textbf{1047} \\
% qf21\_n15 & 15 & 112 & 115 & 208 & 207 & \textbf{197} \\
% dnn\_n16 & 16 & 48 & 384 & 74 & 72 & \textbf{71} \\
% urf1\_278 & 16 & 22267 & 25306 & 46.78k & 46.78k & \textbf{46.77k}  \\
% urf2\_277 & 16 & 8311 & 10062 & 18.29k & 18.29k & \textbf{18.28k}  \\
% cc\_n18 & 18 & 17 & 18 & 27 & 17 & \textbf{17}  \\
ghz\_state\_n23 & 23 & 22 & 22 & 48 & 40 & \textbf{22} \\
% multiplier\_n25 & 25 & 381 & 670 & 671 & 678 & \textbf{659}  \\
swap\_test\_n25 & 25 & 63 & 96 & 120 & 94 & \textbf{89}  \\
% wstate\_n27 & 27 & 26 & 26 & 48 & 60 & \textbf{26}\\
ising\_n50 &  50  &  4  &  98  &  11  &  10  &  \textbf{10}\\
\bottomrule
\end{tabular}
}

\end{threeparttable}
\end{table}

\subsubsection{Scheduling Strategy}

We investigate our methods from two perspectives: gate scheduling and cut-type scheduling. 

\noindent\textbf{Gate scheduling:} According to the results in Table \ref{table: sch-gate-ls}, our method achieves optimal solutions in most benchmarks.
We compare our gate scheduling method with the circuit-order approach in lattice surgery model, where circuit-order denotes scheduling gates based on their appearance in the circuit. Compared with circuit-order, our method optimizes up to 23\% of the execution time.

\begin{table}[ht]
\centering
\renewcommand\arraystretch{1}
\caption{Comparison of different gate scheduling algorithms} \label{table: sch-gate-ls}
\begin{threeparttable}
\resizebox{\linewidth}{!}{
\begin{tabular}{c c c c c c}
\toprule
Circuit name & $n$ & $\alpha$ & $g$ & Circuit-order  &  Ours \\
\midrule
dnn\_n8 & 8 & 48 & 192 & 66 & \textbf{54} \\
grover  &  9  &  110  &  132  &  112  &  114 \\
qpe\_n9 & 9 & 42 & 43 & 42 & \textbf{42} \\
ising\_n10 & 10 & 20 & 90 & 26 & \textbf{20} \\
adder\_n10 & 10 & 55 & 65 & 55 & \textbf{55} \\
QFT\_10 & 10 & 93 & 105 & 93 & \textbf{93} \\
% square\_root.n4 & 11 & 221 & 294 & 225 & \textbf{221} \\
% shor\_a11 & 12 & 6922 & 9248 & 6925 & \textbf{6922} \\
multiply\_n13 & 13 & 23 & 40 & 24 & \textbf{23} \\
% qf21\_n15 & 15 & 112 & 115 & 112 & 112 \\
% dnn\_n16 & 16 & 48 & 384 & 83 & \textbf{67} \\
square\_root\_n18 & 18 & 644 & 898 & 644 & \textbf{644} \\
ghz\_state\_n23 & 23 & 22 & 22 & 22 & \textbf{22} \\
swap\_test\_n25 & 25 & 63 & 96 & 63 & \textbf{63} \\
% multiplier\_n25 & 25 & 381 & 670 & 384 & \textbf{381} \\
% wstate\_n27 & 27 & 28 & 52 & 29 & \textbf{28} \\
ising\_n50 & 50 & 4 & 98 & 9 & \textbf{9} \\

\bottomrule
\end{tabular}
}

\end{threeparttable}

\end{table}

\noindent\textbf{Cut type scheduling:}
As shown in Table \ref{table: sch}, our algorithm outperforms the best baseline strategies on these benchmarks, achieving an average reduction of 25\% and up to a maximum of 50\%. 
We compared the cycle number of our methods with the other two strategies: Time-first and Channel-first. These two strategies determine whether to modify the cut type when dealing with a \cnot~gate with different cut types. The former chooses the operations that make the \cnot~gate complete as soon as possible, while the latter minimizes the channel occupation of this \cnot~gate. 
Our optimization is caused by our strategy of adaptively adjusting the weights of time and channel based on resource conditions, making our strategy perform well in most scenarios.

\begin{table}[ht]
\centering
\renewcommand\arraystretch{1}
\caption{Comparison of different cut type scheduling}
\label{table: sch}
\begin{threeparttable}
\resizebox{\linewidth}{!}{
\begin{tabular}{c c c c c c c}
\toprule
Circuit name & $n$ & $\alpha$ & $g$ & Channel-first& Time-first & Ours \\
\midrule
% qec\_en\_n5 & 5 & 10 & 10 & 10 &10 & 10 \\
dnn\_n8  &  8  &  48  &  192  &  48  &  48  &  \textbf{48} \\
grover & 9 & 110 & 132 & 166 &174 & \textbf{110} \\
qpe\_n9 & 9 & 42 & 43 & 70 &96 & \textbf{42} \\
ising\_n10 & 10 & 20 & 90 & 20 &20 & \textbf{20} \\
adder\_n10 & 10 & 55 & 65 & 78 &88 & \textbf{55} \\
QFT\_10 & 10 & 93 & 105 & 165 &120 & \textbf{93} \\
% BV\_10 & 10 & 5 & 5 & 5 &5 & 5 \\
% sat\_n11 & 11 & 204 & 252 & 336 &387 & \textbf{204} \\
% square\_root.n4 & 11 & 221 & 294 & 379 &402 & \textbf{221} \\
% shor\_a11 & 12 & 6922 & 9248 & 10770 &13862 & \textbf{6922} \\
multiply\_n13 & 13 & 23 & 40 & 35 &41 & \textbf{23} \\
% qf21\_n15 & 15 & 112 & 115 & 197 &265 & \textbf{112} \\
% multiplier\_n15 & 15 & 133 & 222 & 230 &265 & \textbf{133} \\
% dnn\_n16 & 16 & 48 & 384 & 60 &60 & \textbf{48} \\
square\_root\_n18 & 18 & 644 & 898 & 1047 &1117 & \textbf{644} \\
ghz\_state\_n23 &  23  &  22  &  22  &  22  &  22  &  22 \\
swap\_test\_n25 & 25 & 63 & 96 & 89 &102 & \textbf{63} \\
% multiplier\_n25 & 25 & 381 & 670 & 657 &765 & \textbf{381} \\
% BV\_50 & 50 & 27 & 27 & 27 &27 & 27 \\
ising\_n50 & 50 & 4 & 98 & 8 &8 & \textbf{4} \\
% QFT\_50 & 50 & 2363 & 2435 & 4633 &2507 & \textbf{2363} \\
% BV\_100 & 100 & 50 & 50 & 50 &50 & 50 \\
% ising\_n100 & 100 & 4 & 198 & 12 &12 & \textbf{4} \\

\bottomrule
\end{tabular}
}

\end{threeparttable}

\end{table}

\subsubsection{Scalability}
We explore the effectiveness of \ourframework~on various input quantum circuits and chip sizes. Determining the parallelism of a given quantum circuit is challenging, but generating quantum circuits with specified parallelism is feasible. Inspired by QUEKO \cite{tan2020optimality}, we generate 50 random quantum circuits (as a test group) with 49 qubits, 50 depth, and parallelism ranging from 1 to 21. We use the average number of cycles in each group as the result.

\begin{figure}
  \centering
  % 第一个子图
  \begin{subfigure}[b]{0.9\linewidth}
    \centering
    \includegraphics[width=\linewidth]{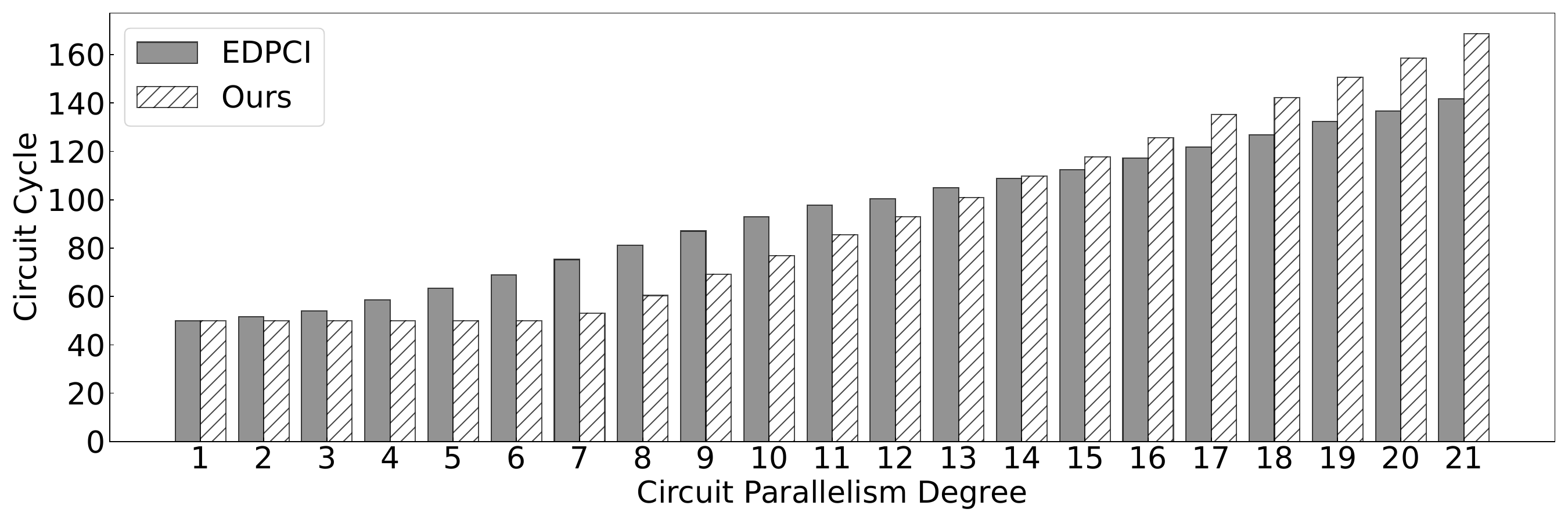}
    \caption{Lattice surgery model}
    \label{paralism_ls}
  \end{subfigure}
  % 第二个子图
  \begin{subfigure}[b]{0.9\linewidth}
    \centering
    \includegraphics[width=\linewidth]{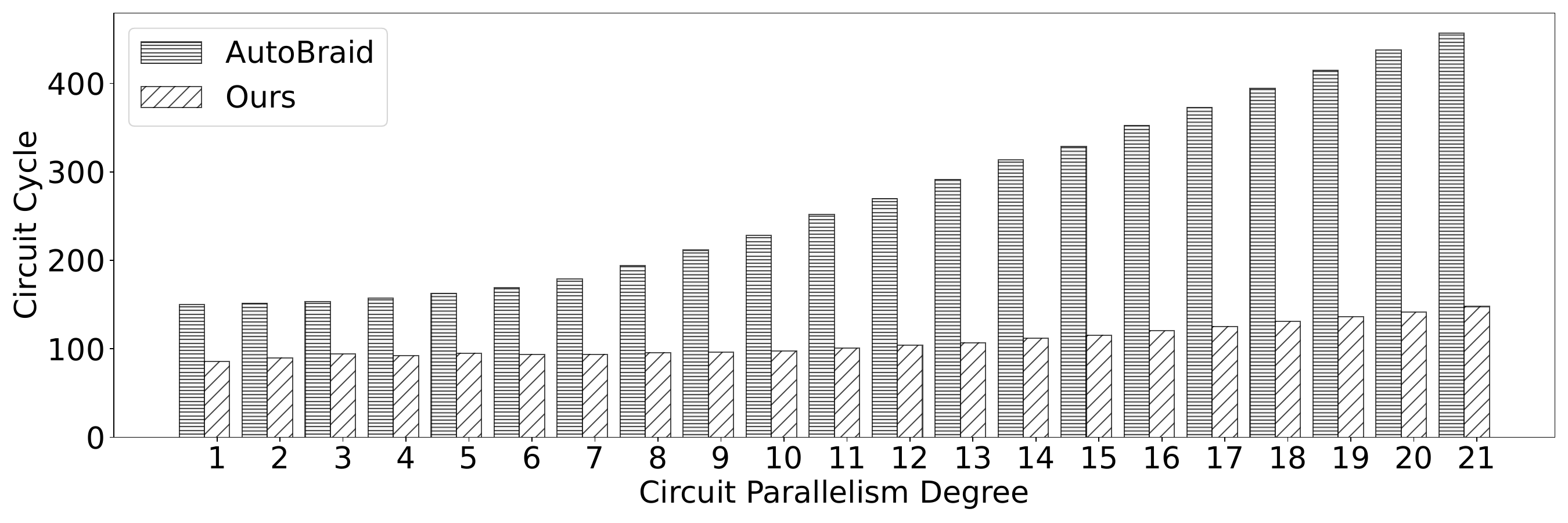}
    \caption{Double defect model}
    \label{paralism_dd}
  \end{subfigure}
  \caption{Effect of circuit parallelism}
  \label{fig:mainfig}
  \vspace{-0.2cm}
\end{figure}
\noindent\textbf{Scalability of Circuit Parallelism Degree:} 
In lattice surgery model, our approach generally outperforms the performance of \PRX~for most circuits, particularly in circuits with parallelism 3 to 13. Our method's performance is slightly less effective for circuits with high $\mathbb{PM}$ than that of \PRX. This is due to our algorithm more likely to get trapped in local optima in these cases.
In double defect model, the optimization ratio increased from 43\% to 62.9\% when \para~increasing from 1 to 21, as shown in Fig. \ref{paralism_dd}. 
This is mainly attributed to our scheduling strategy for the cut type, which effectively leverages the waiting time due to path conflicts. We save significant channel resources by adjusting the cut type when the tile cut types are the same.

\begin{figure}[htbp]
    \centering %图片全局居中  
    \includegraphics[width=1\columnwidth]{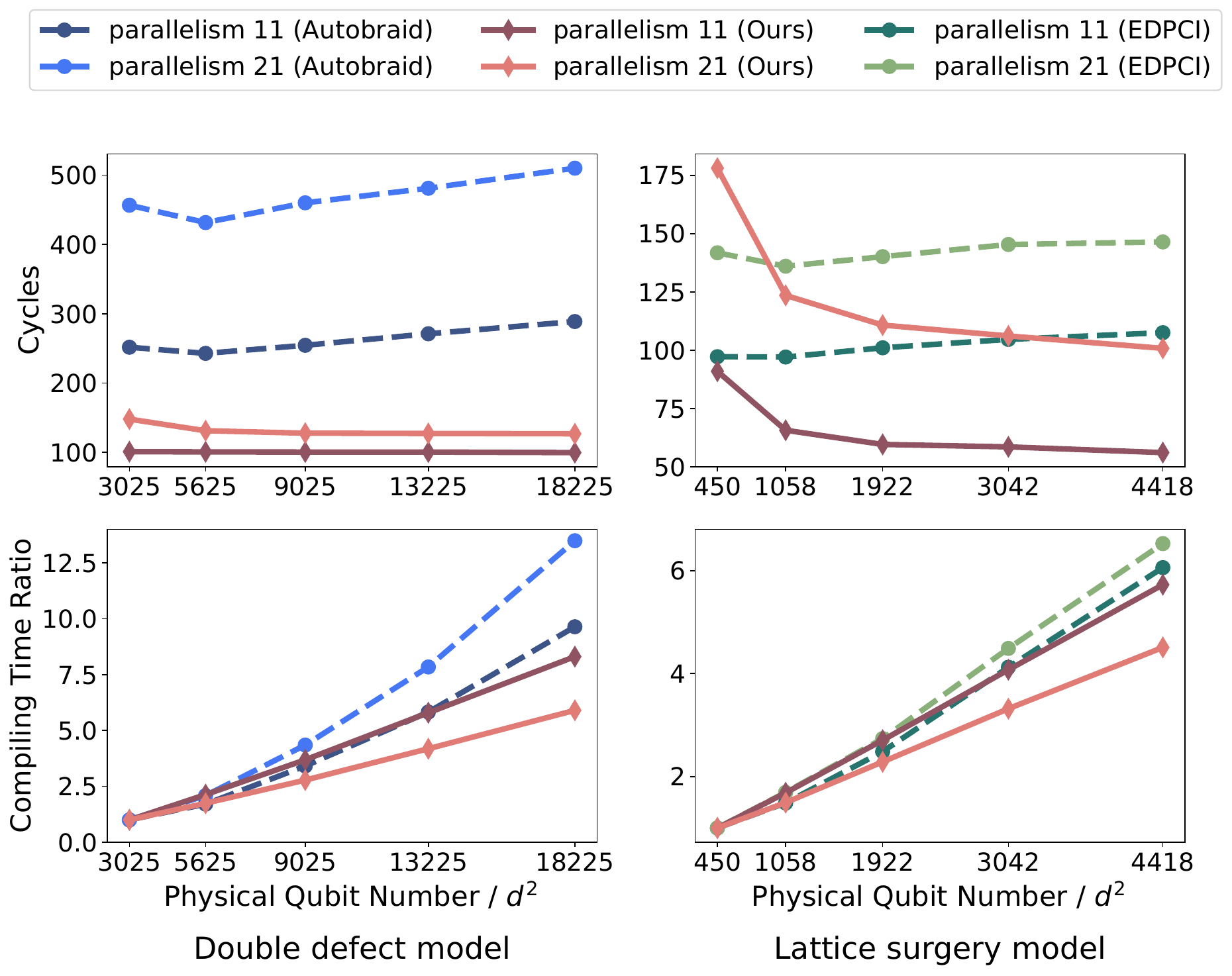}
    % \vspace{0cm} \\
    %并排几个图，就要写几个minipage
    %所有minipage宽度之和要小于1，否则会自动变成竖排
    % \begin{minipage}{0.45\columnwidth}     
    %     \centering %图片局部居中
    %     \includegraphics[width=0.99\columnwidth]{fig/double_defect_chip.pdf} %此时的图片宽度比例是相对于这个minipage的，不是全局
    %     % \vspace{-0.6cm} 
    %     \subcaption{Double defect model}
    %     \label{Bandwidth}
    % \end{minipage}
    % %所有minipage宽度之和要小于1，否则会自动变成竖排
    % \begin{minipage}{0.45\columnwidth}      
    %     \includegraphics[width=0.99\columnwidth]{fig/ChipSize_lattice.pdf}
    %     % \vspace{-0.62cm} 
    %     \subcaption{Lattice surgery model}
    %     \label{Bandwidth_lattice}
    % \end{minipage}
    \caption{Effect of chip size}
    \label{chip_size_impact}
    % \vspace{-0.2cm}
\end{figure}

\noindent\textbf{Scalability of Chip Size: }
Fig. \ref{chip_size_impact} illustrates the trends of \ourframework's performance (cycles) and efficiency (compiling time ratio) as the chip size increases. The compiling time ratio is $\tau_{(i,j)}/\tau_{(i,\min)}$ where $\tau_{(i,\min)}$ is compiling time of circuits with parallelism $i$ at minimum viable chip and $\tau_{(i,j)}$ is compiling time of circuits with parallelism $i$ at chip size $j$. We use this metric to fairly compare the scalability among the three algorithms \ourframework, Autobraid, and \PRX~were programmed in Python, C++, and Julia. The chip is a square with the average bandwidth per channel from 1 to 5. The result demonstrates that our methods' circuit cycles decrease as the chip size increases. The execution time of the circuits with $\mathbb{PM} = 21$ can be decreased by 10.8\% for double defect model and decreased by 30.9\% in lattice surgery model when the average bandwidth of the chip rises from 1 to 2.

%% file: 7-RelatedWork.tex
\section{Related Work}
\label{sec:related work}
Most existing quantum compilers \cite{zulehner2018efficient,siraichi2018qubit,li2019tackling,zhang2021time} focus on the physical qubit level compilation designed for NISQ circuits with 50 to 200 qubits, which is not fault-tolerant. These works focus on converting a logical circuit into a hardware-dependent physical circuit with respect to \cnot~gates applied to physical qubits connected in the hardware. 

Fault-tolerant compilation primarily focuses on architectures based on surface code, as it is the most promising error-correcting code in superconducting quantum computers. 
The fundamental difference between compiling a surface code circuit and a NISQ circuit is separating communication resources (channels) and computational resources (tiles). The resources are software-defined and can be specialized for specific circuits. The execution of CNOT gates is no longer achieved by moving the data to the two physically adjacent physical qubits. Instead, it takes place within the channel, using exclusive access to communication resources.
Depending on the method of constructing logical qubits, surface code can be divided into double defect \cite{fowler2012surface} and lattice surgery \cite{horsman2012surface} with different logical operation implementation strategies.
Double defect employs the braiding technique to perform \cnot gates. Braidflash \cite{javadi2017optimized} abstracts the constraints of \cnot~gates implementation into braiding path disjoint. Autobraid \cite{hua2021autobraid} further discovers the local parallelism pattern and designs a stack-based search algorithm that enables efficient search for as many parallel \cnot~gates as possible.
Lattice surgery is a novel entrant in surface code approaches, employing a reduced number of physical qubits for encoding a logical qubit. It utilizes ZZ measurements for \cnot~gate \cite{litinski2019game}. 
%However, the physical adjacency requirement for measurements limits its effectiveness. As a result, executing long-range \cnot~gates is time-consuming. For example, a \cnot~gate with a distance of $k$ requires $k\times d$ surface code cycles, where $d$ is the code distance.
\PRX~\cite{beverland2022surface} achieves long-range \cnot~gates by utilizing ancilla tiles to construct Bell states. This approach requires a fourfold increase in physical qubits but enables the completion of \cnot~gates at arbitrary distances within $2d$ surface code cycles. However, this approach does not account for the impact of initial mapping. Disregarding the circuit communication requirements with a trivial initial mapping results in a paradoxical situation where the circuit's performance worsens as chip resources increase. 

Other works on fault-tolerant quantum compilation have focused on synergy with chip characteristics. Wu \etal \cite{wu2022synthesis} proposes a lattice encoding method for superconducting chips, which adapts the various chip structures to the surface code's 2D lattice. Previous works \cite{mcewen2023relaxing,chamberland2020topological} involve adapting the surface code to hexagonal chips, reducing chip connectivity to improve the accuracy of physical qubits. Some efforts \cite{mcewen2023relaxing} focus on utilizing operations with lower error rates during the compilation process to enhance circuit accuracy. Preskill \etal \cite{pattison2023hierarchical}, on the other hand, centers on concatenating surface code and high-rate code, like quantum LDPC encoding, to address the challenges of low code rate and limited scalability in surface code implementations.

%% file: 8-Conclusion.tex
\section{Conclusion}
\label{sec:conclusion}
In this paper, we study the surface code mapping and scheduling problem for the lattice surgery and double defect models. We formalize the problems in both models and establish the problem's complexity, particularly highlighting challenges in the double defect model.
We introduce \para~and \chipcapacity~to quantitatively analyze quantum circuits and quantum chips.
Our mapping and scheduling methods, named \ourframework, feature algorithms for scenarios with sufficient and limited qubit resources. Extensive evaluations show that \ourframework~provides significant reduction over the state-of-the-art approaches by reducing the execution time by 33.3\% to 67.3\% for double defect model and reducing by up to 13.9\% for lattice surgery model.

\textbf{Limitation and future work}: As \para~and \chipcapacity~are critical parameters for our mapping and scheduling methods, we still lack effective algorithms to obtain accurate results. In addition, our cost function to determine the importance of the current gate shows less effectiveness for high-parallel circuits compared to circuits with lower parallelism. Our investigation anticipates dynamic transforming strategies modifying mappings during the transforming process. Moreover, the complexity of scheduling problems and the bounds of chip communication capacity are still open problems.

%% file: A.tex
\section{Proof of Theorem 1}
\label{appendix:A}
\begin{figure}[htbp]
	\centering  %图片全局居中
	\includegraphics[width=0.95\linewidth]{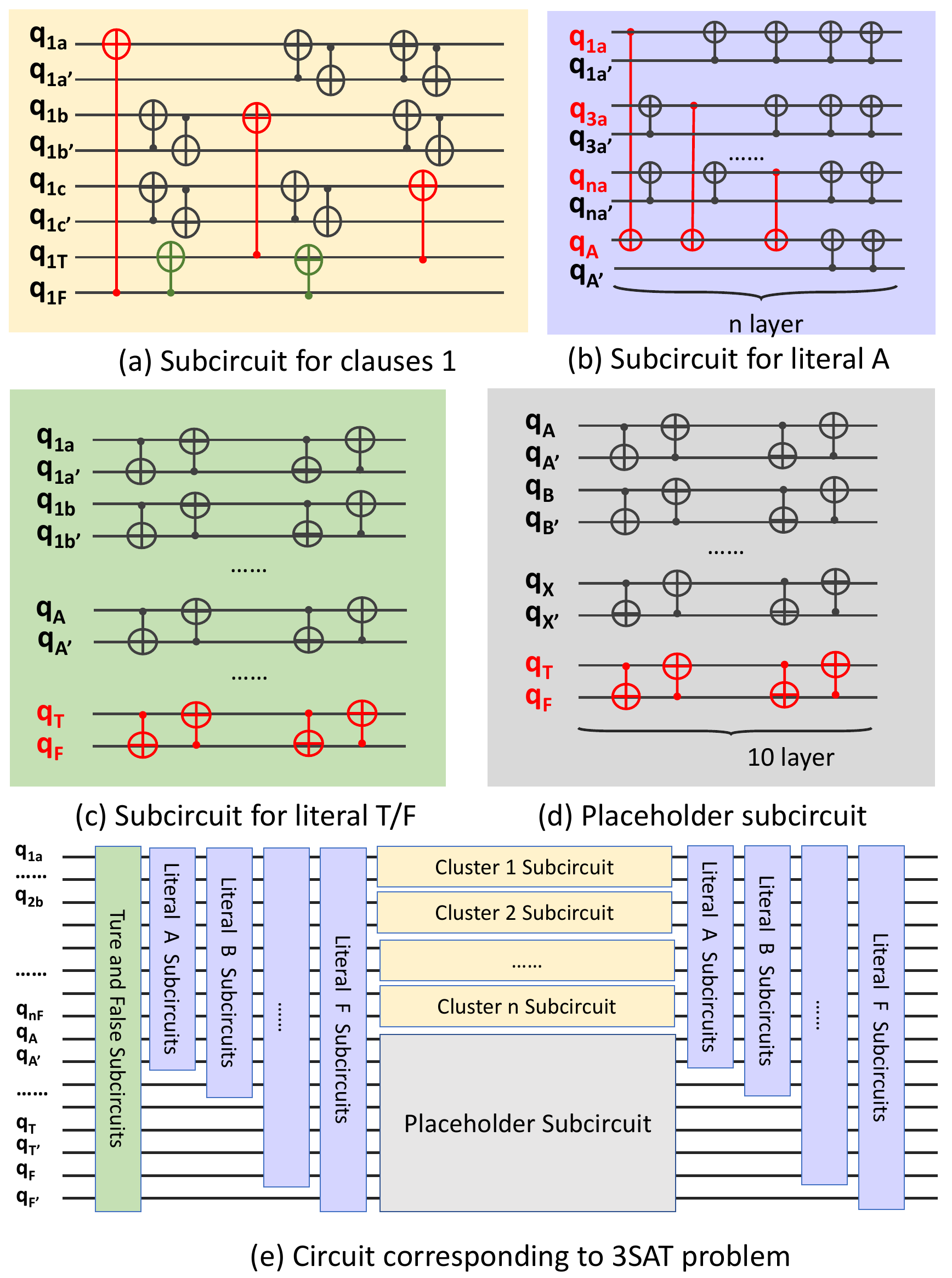}
	\caption{Quantum circuit construction for an $n$-clause 3-SAT Problem}
	\label{npc}
\end{figure}

We prove that any instance of a 3-SAT problem can be reduced to an instance of a surface code initialization problem in polynomial time. We can construct the corresponding quantum circuit for any $n$-clause 3-SAT problem that the 3-SAT problem can be satisfied if and only if an initialization exists to execute this circuit no more than $10+3n$ cycles. Here, we assume that the bandwidth of the channel is sufficient. The quantum circuit, as shown in Fig.\ref{npc}e, is constructed in the following way:

For each three-literal clauses $C_i$ construct a sub-circuit with $8$ qubits, $q_{ia},q_{ib},q_{ic}$ represent three literals $a,b,c$ in the clause, $q_{ia'},q_{ib'},q_{ic'}$ are the ancilla qubits respectively, and $q_{iT}, q_{iF}$ represent the logical qubits initialized into \xcut~tile and \zcut~tile. If the clause's first literal $a$ is positive, we add a \cnot~gate between $q_{1a}$ and $q_{1T}$, otherwise between $q_{1a}$ and $q_{1F}$. Then we add a \cnot~gate between $q_{1T}$ and $q_{1F}$. After that we add two ancilla \cnot~gates between $q_{1b}$ and $q_{1b'}$, as well as $q_{1c}$ and $q_{1c'}$. For the second and third literals $b$ and $c$, the circuit is constructed in the same way as above. For example, the sub-circuits corresponding to clauses $(a \lor \lnot b \lor  c)$ are shown in Fig.\ref{npc}a. If the clause is true, a tail mapping exists, allowing the depth of this sub-circuit to be no more than 10. The cut type of each tile is mapped one-to-one with the true and false values of this literal. Here, the black gates are used for placeholders so that the tiles do not have time to change their tile type within 10 cycles. 

Each three-literal clause generates a corresponding sub-circuit, which we connect in parallel, and the depth of it is no more than $10$ if and only if all the clauses are True. Next, we must ensure that the same literal in different clauses corresponds to the same cut type. This is achieved through sub-circuit in Fig.\ref{npc}b. We declare an ideal literal and let the literal in different clauses perform \cnot~gates with it. The circuit can achieve its shortest depth only if they are of different types to the ideal literal. Here, the black gate is used for placeholder operations if a tile modifies its tile type and to supplement the circuit so that the shortest depths corresponding to different literals are $n$. For the ideal True and False, we require an additional n-depth sub-circuit in Fig.\ref{npc}c that makes their cut types different. Sub-circuit Fig.\ref{npc}d ensures that the ideal literal does not modify its cut type while waiting for the clauses sub-circuit to execute.

%最后将他们组合起来，这个量子线路的深度小于等于$10+3n$，当且仅当
% Combining these sub-circuits, as depicted in Fig.\ref{npc}e, forms the corresponding quantum circuit. There is an initial mapping of this quantum circuit resulting in cycles no more than $3n+10$ if and only if the 3SAT problem has a solution.

